\documentclass[11pt,a4paper]{article}
\usepackage{float}
\usepackage{epsfig}
\usepackage{multirow}
\usepackage[cp1252,utf8]{inputenc}
\usepackage{hyperref}
\usepackage{graphicx} 
\usepackage{amsmath}
\usepackage{amsfonts}
\usepackage{amssymb}
\usepackage{bigints} 
\usepackage[top=2cm, bottom=2.5cm, left=1.5cm, right=1.5cm]{geometry}
\usepackage{authblk}
\usepackage{changepage} 
\usepackage{multicol}
\usepackage{afterpage}
\usepackage[normalem]{ulem}
\usepackage{placeins}

\usepackage{ragged2e}
\usepackage{graphicx,wrapfig}
\usepackage{bm}
\usepackage{hyperref}
\usepackage{mathrsfs}
\usepackage{subfig}
\usepackage{breqn}
\usepackage{caption}

\usepackage[english]{babel}

\title{\bf Stability, quasinormal modes in a charged black hole in perfect fluid dark matter}
\author{\bf  Anish Das \thanks{anishdas1995@bose.res.in}}
\author{\bf  Anirban Roy Chowdhury \thanks{iamanirban.rkmvc@gmail.com}}
\author{\bf Sunandan Gangopadhyay \thanks{sunandan.gangopadhyay@bose.res.in}}

\affil{\textit{Department of Astrophysics and High Energy physics,\linebreak
S.N.~Bose National Centre for Basic Sciences,}\linebreak
\textit{JD Block, Sector-III, Salt Lake, Kolkata 700106, India}}

\date{}

\begin{document}
\maketitle

\begin{abstract}

\noindent In this work, we study time-like and null geodesics in a charged black hole background immersed in perfect fluid dark matter (PFDM). Using the condition for circular geodesics, we evaluate the energy ($E$) and angular momentum ($L$) in terms of the radius ($r_c$) of the circular orbits. The existence and finite-ness of $E$ and $L$ constrain the possible range of PFDM parameter ($\chi$) and the radius of the circular orbit ($r_c$). We then use the Lyapunov exponent ($\lambda$) to study the stability of the geodesics. Then we analyze the critical exponent ($\gamma$) useful for determining the possibility of detection of gravitational wave signals. After that, we study the perturbation due to a massless scalar field in such a background and calculate the quasinrmal mode (QNM) frequencies and their dependence on PFDM parameter $\chi$ and black hole charge $Q$. Also, we compare the obtained QNM frequencies both in the exact case and in the eikonal limit. We also calculate the quality factor of the oscillating system and study its dependence on $\chi$ and $Q$. Finally, we evaluate the black hole shadow radius $R_s$ and graphically observe the effect of $\chi$ and $Q$ on it.

\end{abstract}

\maketitle
\section{Introduction}\label{sec0}
The theory of general relativity proposed by Einstein has led to the predictions of compact objects such as black holes whose existence has been verified recently \cite{1}-\cite{3}. Also, Einstein's theory predicts that spacetime is dynamic and gets affected by the presence of any massive object. The presence of compact objects like black holes drastically alters the spacetime fabric in their vicinity. This effect can be observed in the trajectories of particles moving in the black hole background. Any object in an arbitrary spacetime in absence of additional forces follows the geodesics. The knowledge of geodesics helps us gain information about the spacetime background.  

\noindent Again, the geodesics around black holes comprise different closed and open orbits depending on their position with respect to the black holes. The orbits can be both stable and unstable. The stability/instability can be understood by understanding the potential ($V$) that the particles encounter. Another way to study the stability of geodesics is by using the Lyapunov exponents ($\lambda$) \cite{4},\cite{5}. The Lyapunov exponents are very powerful mathematical objects useful for studying a system having chaotic dynamics generally encountered in non-linear systems. We can also use them to study the stability of particle trajectories. Lyapunov exponents $\lambda$ are mathematically defined as the average rate at which two nearby geodesics or geodesic congruences can converge or diverge \cite{6}. The principal Lyapunov exponent can be expressed in terms of the second derivative of the potential evaluated at the extremum point of the potential. 

\noindent In general, black holes are surrounded by matter (accretion disks) which can result in perturbing the spacetime structure. This perturbation influences the geodesics in the black hole spacetime and the information of the perturbation is carried by the geodesics. Again, whether such perturbations will grow or not decides the stability of the black holes against such perturbation. Due to such perturbation, the black hole spacetime starts to oscillate with a particular frequency termed `quasi-normal frequency' coined by Press \cite{7}. The term quasi-normal is used because the oscillations can either grow or decay depending on the black hole's stability. The detection of these frequencies helps us get information on the space-time parameters and is independent of the type of perturbation. The study of quasinormal modes is also important in the context of the AdS/CFT correspondence \cite{ff1}. We can study the equilibrium and non-equilibrium properties of strongly coupled thermal gauge theories by computing the QNMs of its gravity dual \cite{ff2}. The spectrum of the quasinormal modes of the dual gravitational background gives us the poles of the retarded correlators in the filed theory side \cite{ff3,ff4}. Investigations of the quasinormal modes is also important in the context of the astrophysical black holes and gravitational wave astronomy \cite{ff5}. Black hole quasinormal modes can be used to predict the mass, angular momentum and other important properties of the astrophysical black holes \cite{ff5}-\cite{ff8}. A lot of studies have been done on black hole QNMs \cite{7a}-\cite{33} along with review works \cite{34}-\cite{37}. 

\noindent Again, the universe is composed mostly of dark matter and dark energy. Dark matter dominates the matter content of the universe. Many different models for dark matter have been proposed like cold dark matter, axions, neutralinos, etc \cite{38}. But these models have failed to explain certain aspects like the too-big-to-fail problem \cite{39}, \cite{39a}, missing satellite problem \cite{40}, etc. The failures of all those models led to the interest in newer models like the perfect fluid dark matter (PFDM) model proposed by Kiselev \cite{41}, \cite{42}. Further works in this  PFDM model have been done in \cite{43}-\cite{49}. Also, quasinormal mode frequencies in the presence of PFDM have been done previously in \cite{23},\cite{23aa}. The consideration of dark matter around the black holes is realistic since dark matter is supposed to pervade all around the galaxies. The model of our choice namely the perfect fluid dark matter (PFDM) has gained popularity in recent years. Also, the consideration of charge $Q$ is for completeness and we refrain from adding spin $a$ to the black hole for simplicity. Theoretically, the effect of dark matter must be present in all observable features of a black hole and hence considering it to study different aspects of a black hole is worthwhile.

\noindent In this work, we are interested in studying the stability of geodesics around black holes by using the concept of Lyapunov exponents $\lambda$. The stability of geodesics in general for any black hole spacetime is determined by the potential and the corresponding conditions imposed onto it. We will find that the Lyapunov exponent is related to the second derivative of the potential $V''$. Also, we use different times (proper and coordinate time) to determine $\lambda$ to check whether they are coordinate-dependent or not. Then we calculate the critical exponent $\gamma$ \cite{49a}, \cite{49b} which helps us get an idea of the timescale at which gravitational waves are possible for detection.

\noindent Again, we are interested in studying the scalar perturbations which result in the production of quasinormal frequencies. The quasinormal frequencies are independent of the type of perturbation and depend only on the black hole parameters. Here, we assume a black hole immersed in a dark matter background to analyze the dependence of  black hole QNMs on PFDM parameter $\chi$. Also in the eikonal limit \cite{50}, \cite{51} we determine the QNMs and compare them with the exact ones. Then we determine the quality factor ($QF$) \cite{52}, \cite{22a} and how it depends on $\chi$ and $Q$. Then we study the shadow of charged black hole \cite{k1}, \cite{k2} in PFDM and its dependence on $\chi$ and $Q$.

\noindent The paper is organized as follows. In section \ref{sec0} we provide the background and brief literature on PFDM, QNMs, and black hole stability. In section \ref{sec1}, we provide the mathematical formulation for studying black hole stability along with the idea of Lyapunov exponents $\lambda$ and critical exponent $\gamma$. After that in section \ref{sec2} we overview the system of charged black holes immersed in PFDM and study the geodesics along with the relevant physical quantities quantifying the stability of geodesics in black hole spacetime. Then in section \ref{sec3} we study the quasinormal modes due to massless scalar perturbation on the background of interest. We study and compare the exact and eikonal approximated results. Later in section \ref{sec4} we study the black hole shadow and how it gets affected by PFDM parameter $\chi$ and black hole charge $Q$. Finally, we summarize in section \ref{sec5}. We have worked in units of $c=G=1$. The signature of our metric is (-,+,+,+).

\section{Equatorial geodesics in general static spherically symmetric spacetime}\label{sec1}
In this work, we are interested in studying the equatorial geodesics of a general static spherically symmetric spacetime. The knowledge of these geodesics will help us determine the stability of the spacetime against external perturbations. To carry out our analysis, we consider a general static spherically symmetric metric in (3+1)-dimensions. We assume the metric to be
\begin{equation}\label{1}
    ds^2=-f(r)dt^2 + \frac{dr^2}{f(r)} + r^2 d\theta^2 + r^2 \sin^2 \theta d\phi^2
\end{equation}
where $f(r)$ corresponds to the lapse function and is a function of $r$ only. We derive the expressions of geodesics in terms of $f(r)$.

\noindent Since our system is spherically symmetric, all planes are identical and hence for simplicity, we calculate the geodesics in the equatorial plane. The condition for the equatorial plane is $\theta=\frac{\pi}{2}$ which results into $\dot{\theta}=0$. The metric is independent of $t$ and $\phi$  coordinates, so the corresponding generalized momenta are conserved. This can be checked by using a Lagrangian of the form $\mathcal{L}=\frac{1}{2}g_{\mu \nu}\dot{x}^{\mu}\dot{x}^{\nu}$ and the Lagrange's equation of motion which takes the form
\begin{equation}\label{2}
    \frac{d}{d\zeta}\Big(\frac{\partial \mathcal{L}}{\partial \dot{x}^{\mu}}\Big)=\frac{\partial \mathcal{L}}{\partial x^{\mu}}~.
\end{equation}
The generalised momenta is defined as $p_{\mu}=\Big(\frac{\partial \mathcal{L}}{\partial \dot{x}^{\mu}}\Big)$ which for $t$ and $\phi$ gives $p_t =$ constant $= -E$ and $p_{\phi}=$ constant $=L$ (Since the metric is independent of $t$ and $\phi$ as mentioned above and also $x^{\mu}$ and $\dot{x}^{\mu}$ are independent of each other). The geodesics for $t$ and $\phi$ take the form
\begin{eqnarray}\label{3}
    \frac{dt}{d\zeta}=\frac{E}{f(r)}~~;~~ \frac{d\phi}{d\zeta}=\frac{L}{r^2}
\end{eqnarray}
where $\zeta$ parametrizes the geodesic trajectory. The radial geodesic equation can be obtained using the normalization condition
\begin{equation}\label{4}
    g_{\mu \nu}\dot{x}^{\mu}\dot{x}^{\nu}=\beta~
\end{equation}
with, $\beta=-1,0,1$ for time-like, null, and space-like geodesics respectively. Using the metric coefficients (eq.\eqref{1}) and the geodesic equations (eq.\eqref{3})  in eq.\eqref{4}, we obtain the radial geodesic equation as
\begin{equation}\label{5}
    \dot{r}^2 = E^2 - f(r)\Big(\frac{L}{r^2} - \beta\Big)= V_{r}~.
\end{equation}
Here, $V_r$ corresponds to the effective radial potential. We are interested in studying circular geodesics which are subject to the condition 
\begin{equation}
    \dot{r}^2\Big|_{r=r_c} = \Big(\dot{r}^2\Big)^{'}\Big|_{r=r_c}=0~.
\end{equation}
Here, prime ($'$) corresponds to the derivative with respect to $r$ and $r_c$ corresponds to the radius of circular geodesics. Since only time-like and null geodesics have a physical existence, so we study them case-wise.
\subsection{Time-like geodesics}
Time-like geodesics correspond to the trajectory of massive particles in any spacetime background. In this case, we have constant quantities $E$ and $L$ corresponding to momentum conservation along $t$ and $\phi$ directions. Here, $E$ and $L$ correspond to the energy and angular momentum per unit mass of the particles moving along time-like geodesics. Since $\beta=-1$ for time-like geodesics, we obtain the radial geodesic equation as
\begin{equation}\label{7a}
    \dot{r}^2 = E^2 - f(r)\Big(\frac{L^2}{r^2} + 1\Big)= \overline{V}_{r}~.
\end{equation}
Here, $\overline{V}_{r}$ corresponds to the effective potential encountered by the massive particles. The condition for circular geodesics are
\begin{equation}
    \dot{r}^2\Big|_{r=r_0} = \Big(\dot{r}^2\Big)^{'}\Big|_{r=r_0}=0 
\end{equation}
with $r_0$ being the radius of circular time-like geodesics. As we can see, here we have two conditions (constraints) whereas three undetermined quantities $E, L$ and $r_0$. So we need to express any two of the physical quantities in terms of the other. We wish to express $E$ and $L$ in terms of $r_0$. The conditions result in the following equations
\begin{eqnarray}\label{9}
    E_0 ^2 = f(r_0)\Big(\frac{L_0 ^2}{r_0 ^2} + 1\Big)
\end{eqnarray}
\begin{eqnarray}\label{10}
    2L_0 ^2 \frac{f(r_0)}{r_0 ^3}=f'(r_0)\Big(\frac{L_0 ^2}{r_0 ^2} + 1\Big)~.
\end{eqnarray}
Solving eq.(10), we obtain $L_0 ^2$. Using the value of $L_0 ^2$ in eq.(9), we get the expression for $E_0 ^2$. The expressions for $E_0 ^2$ and $L_0 ^2$ take the form \cite{6}
\begin{equation}\label{6}
    E_0 ^2 = \frac{2 f^2 (r_0)}{2f(r_0)-r_0 f'(r_0)}
\end{equation}
\begin{equation}\label{a6}
    L_0 ^2 = \frac{ r_0 ^3 f' (r_0)}{2f(r_0)-r_0 f'(r_0)}~.
\end{equation}
Here, $E_0$ and $L_0$ give the energy and momentum per unit mass of particles moving along circular geodesics of radius $r_0$. The ratio of angular momentum $L_0$  and energy $E_0$ takes the form
\begin{equation}\label{10}
    \Big(\frac{L_0}{E_0}\Big)^2 = \frac{r_0 ^3 f'(r_0)}{2f^2 (r_0)}~.
\end{equation}
In eq.\eqref{6}, we find the expressions for the square of the quantities $E_0$ and $L_0$. For $E_0$ to be real and finite, we must have the denominator of eq.\eqref{6} to be positive, that is
\begin{equation}
    \Big(2f(r_0)-r_0 f'(r_0)\Big)>0
\end{equation}
and for $L_0$ to be real and finite, we must simultaneously have both the numerator and denominator of eq.\eqref{a6} to be positive, that is
\begin{equation}
 r_0 ^3 f' (r_0) > 0~;~\Big(2f(r_0)-r_0 f'(r_0)\Big)>0~.
\end{equation}
\subsection{Null geodesics}
Null geodesics correspond to trajectories of massless (null) particles in any space-time background. In the case of null geodesics, we have $\beta=0$ which results in the radial geodesic equation in the equatorial plane as
\begin{equation}\label{5a}
    \dot{r}^2 = E^2 - f(r)\frac{L^2}{r^2} = \widetilde{V}_{r}~.
\end{equation}
with $\widetilde{V}_{r}$ giving the effective potential. The condition for circular geodesics results in the determination of radius $r_p$ of null geodesics through the equation
\begin{equation}\label{7}
     2f(r_p)-r_p f'(r_p)=0~.
\end{equation}
Also, it helps us determine the ratio of angular momentum $L_p$ and energy $E_p$ of the massless particles in terms of $r_p$ 
\begin{equation}\label{81}
\Big(\frac{L_p}{E_p}\Big)^2 = \frac{r_p ^3 f'(r_p)}{2f^2 (r_p)} = \frac{r_p ^2}{f(r_p)}~.  
\end{equation}
Here, $E_p$ and $L_p$ give the energy and momentum of null particles moving along circular geodesics. From eq.(s)\eqref{10} and \eqref{81} we find that the ratio of $L$ and $E$ are same for both massive and massless particles.\\

\subsection{Brief review on Lyapunov exponents}
Lyapunov exponents measure the rate of divergence or convergence of nearby trajectories in a dynamical
system.  In a dynamical system,
the state of the system at a given time is described by a set of variables that can be represented as a
point in a multi-dimensional space. Trajectories in this space represent the evolution of the system over
time, and nearby trajectories may either converge or diverge depending on the nature of the system.

\noindent If the Lyapunov exponent is positive, the system is said to be chaotic, meaning that
small perturbations grow exponentially and the behaviour of the system becomes unpredictable over time.
If the Lyapunov exponent is negative, the system is said to be stable, meaning that small perturbations
converge to a fixed point or periodic orbit. The vanishing of the Lyapunov exponent means that the rate of
divergence or convergence of nearby trajectories approaches zero, indicating that the system is marginally
stable. This means that small perturbations in the system will not grow or decay over time.

\noindent The equation of a dynamical system takes the form \cite{69}
\begin{equation}\label{14}
    \frac{dx}{dt}=F(x)
\end{equation}
where, $x(t)$ denotes the trajectory of the system, and the evolution of the system is dictated by the function $F(x)$. We perturb the system slightly which takes $x(t) \to x(t) + \delta x(t)$. Using this in eq.\eqref{14}, we have (upto linear order)
\begin{equation}\label{15}
    \frac{d(\delta x)}{dt}=\frac{\partial F}{\partial x}\Bigg|_{x}\delta x~.
\end{equation}
Generalizing this to higher dimensions, we obtain
\begin{equation}\label{16}
    \frac{d(\delta X_i (t))}{dt}=\frac{\partial F_i (X_j)}{\partial X_j}\Bigg|_{X_i}\delta X_j (t)= K_{ij} (t) \delta X_j (t)~~;~~i,j=1,....,N
\end{equation}
with $K_{ij} (t)$ being the \textit{linear stability matrix} \cite{50}. From the knowledge of the one-dimensional solution, the solution can be extended to higher dimensions as 
\begin{equation}
   \delta X_i (t) = L_{ij}(t) \delta X_j (0) 
\end{equation}
where $L_{ij} (t)$ is the evolution matrix \cite{50} determining the evolution of the perturbation on the system with the property $L_{ii}(0)=\delta_{ii}$. The eigenvalues of the linear stability matrix correspond to the Lyapunov exponents ($\lambda$) which are responsible for determining the stability of the system. The solution of eq.\eqref{16} determines $\lambda$ as
\begin{equation}
\lambda = \lim_{t \to \infty} \frac{1}{t} ln \Bigg(\frac{\delta X_i (t)}{\delta X_i (0)}\Bigg)=\lim_{t \to \infty} \frac{1}{t}ln \Bigg(\frac{L_{ii} (t)}{L_{ii} (0)}\Bigg)~.
\end{equation}
We wish to calculate the Lyapunov exponent ($\lambda$) in order to study the stability of the system. To do so, we perturb the geodesic equations to obtain equations in linearised form as eq.\eqref{15}. The eigenvalues of the matrix $K_{ij}$ give the Lyapunov exponents. 

\noindent As mentioned above, we are interested in circular orbits that lie in the equatorial plane. Thus we are concerned with the radial geodesic equation. For this, we will work in the phase space ($p_r, r$). The analysis can be carried out both for proper time $\tau$ and coordinate time $t$. 

\noindent We start with the Euler-Lagrange equation which takes the form
\begin{equation}\label{19}
    \frac{d p_r}{d \tau} = \frac{\partial \mathcal{L}}{\partial r}
\end{equation}
representing the evolution equation of $p_r$. The equation for $r$ takes the form
\begin{equation}\label{20}
    \frac{dr}{d \tau}= \frac{p_r}{g_{rr}}
\end{equation}
using Lagrange's equation of motion. The system represented by the combined eq.(s) \eqref{19}, \eqref{20} are perturbed with $r \to r + \delta r$ and $p_r \to p_r + \delta p_r$. The perturbed equation takes the form
\begin{equation}\label{21}
    \frac{d (\delta p_r)}{d \tau}= \frac{d}{dr}\Big(\frac{\partial \mathcal{L}}{\partial r}\Big)\delta r~~;~~ \frac{d (\delta r)}{d \tau}= \frac{1}{g_{rr}} \delta p_r~.
\end{equation}
The reason for the above results is that in eq.\eqref{19}, the Lagrangian $\mathcal{L}$ is independent of $p_r$ and in eq.\eqref{20} we have $\dot{r}=\frac{p_r}{g_{rr}}$ which in dependent of $r$. The above equations can be written in matrix form as
\begin{equation}
\frac{d}{d \tau}\begin{pmatrix}\delta p_r \\ \delta r  \end{pmatrix} =    \begin{pmatrix}
  0 & \frac{d}{dr}\Big(\frac{\partial \mathcal{L}}{\partial r}\Big)\\ 
  \frac{1}{g_{rr}}  & 0
\end{pmatrix} \begin{pmatrix} \delta p_r \\ \delta r  \end{pmatrix}~.
\end{equation}
The eigenvalues of the matrix 
\begin{equation}
    K_{ij}= \begin{pmatrix}
  0 & \frac{d}{dr}\Big(\frac{\partial \mathcal{L}}{\partial r}\Big)\\ 
  \frac{1}{g_{rr}}  & 0
\end{pmatrix} 
\end{equation}
evaluated at the circular orbit $r_c$ subject to the condition $\dot{r}^2\Big|_{r=r_c} = \Big(\dot{r}^2\Big)^{'}\Big|_{r=r_c}=0$ gives the Lyapunov exponent 
\begin{equation}
    \lambda^2 = \frac{1}{g_{rr}}\frac{d}{dr}\Big(\frac{\partial \mathcal{L}}{\partial r}\Big)\Bigg |_{r=r_c}~.
\end{equation}
Using the condition of circular geodesics and after some simplification, the proper time Lyapunov exponent reads \cite{50}
\begin{equation}\label{L1}
    \lambda_p = \pm \sqrt{\frac{V ''_r}{2}}\Bigg|_{r=r_c}~.
\end{equation}
On the other hand, the coordinate time Lyapunov exponent takes the form \cite{50}
\begin{equation}\label{L2}
    \lambda_c = \pm \sqrt{\frac{V ''_r}{2 \dot{t}^2}}\Bigg|_{r=r_c}~.
\end{equation}

\subsection{Critical exponent}
Another quantity of interest is the critical exponent $\gamma$ defined
as the ratio of the Lyapunov timescale or instability timescale $T_{\lambda}$ to the orbital timescale $T_{\Omega}$ given as \cite{49a}-\cite{50}
\begin{eqnarray}\nonumber
    \gamma &=& \frac{Lyapunov~~ timescale}{Orbital ~~timescale}\\ \nonumber
    &=&\frac{T_{\lambda}}{T_{\Omega}}\\ 
    &=&\frac{\Omega}{2 \pi \lambda} 
\end{eqnarray}
with $T_{\lambda}=\frac{1}{\lambda}$ and $T_{\Omega}=\frac{2\pi}{\Omega}$, with the orbital velocity $\Omega$ given as  $\Omega=\frac{d\phi}{dt}$. The critical exponents corresponding to proper and coordinate time take the form
\begin{equation}\label{032}
    \gamma_p=\frac{\Omega}{2 \pi \lambda_p}~~;~~\gamma_c=\frac{\Omega}{2 \pi \lambda_c}~~.
\end{equation}
The critical exponent $\gamma$ gives an idea of the detectability of gravitational wave signals. The particles moving in circular orbits around the black hole produce gravitational waveforms. This waveform reaches the observer in every $T_{\Omega}$ time interval. If the particle in the orbit is perturbed, then the gravitational waveform due to perturbation reaches the observer after a time interval of $T_{\lambda}$ dictated by the Lyapunov exponent $\lambda$. If the perturbed signal reaches the observer within the time interval $T_{\Omega}$, then only the perturbation and thereby the corresponding gravitational signal can be detected by the observer. If the signal produced due to perturbation reaches the observer after a time interval $T_{\Omega}$, then the observer cannot distinguish the signals, which one is due to circular motion and which one is due to perturbation. Thus the requirement for the detection of gravitational signals produced due to perturbation is $T_{\lambda} < T_{\Omega}$ \cite{49a}-\cite{50}.

\section{Charged black hole in perfect fluid dark matter }\label{sec2}
The analysis carried out above is valid for any arbitrary static spherically symmetric metric of the form eq.\eqref{1}. Here we want to explicitly study a system where a \textit{charged black hole is immersed in a background of perfect fluid dark matter (PFDM)} \cite{47}-\cite{49}. The action and the corresponding equation of motion for the system have the form \cite{47}-\cite{49} (with $c=G=1$)
{\begin{eqnarray}
S=\int d^4 x\sqrt{-g}\Big(\frac{R}{16\pi } - \frac{1}{4}F^{\mu \nu}F_{\mu \nu} +\mathcal{L}_{DM}\Big) 
\end{eqnarray}
\begin{eqnarray}
R_{\mu \nu}-\frac{1}{2}g_{\mu \nu}R=8\pi  \Big( T_{\mu\nu}^{M}-T_{\mu\nu}^{DM}\Big)~.
\end{eqnarray}
Here $R$ and $ R_{\mu \nu}$ are the Ricci scalar and Ricci tensor respectively. $F_{\mu \nu}$ is the electromagnetic field strength tensor related to 4-vector potential $A_{\mu}$, $F_{\mu \nu}=\partial_{\mu} A_{\nu} - \partial_{\nu} A_{\mu}$. Also, $\mathcal{L}_{DM}$ gives the Lagrangian density of the PFDM. $T_{\mu\nu}^{M}$ and $T_{\mu\nu}^{DM}$ give the energy-momentum tensor for the electromagnetic field and the dark matter respectively. The energy-momentum tensors take the form 
\begin{eqnarray}
\Big(T^{\mu}_{~\nu}\Big)^{M}=\frac{Q^2}{8\pi r^4}diag\left(-1, -1 , 1, 1\right); \Big(T^{\mu}_{~\nu}\Big)^{DM}=diag \Big(-\rho, P_r, P_{\theta}, P_{\phi} \Big)
\end{eqnarray}
with $P_r = -\rho$, $P_{\theta}= P_{\phi}=P$. The equation of state for PFDM is $\frac{P}{\rho}=\frac{1}{2}$ \cite{48}.

\noindent To obtain the desired metric, we need to solve the Einstein field equations. The first step is to assume an ansatz metric of the form 
\begin{equation}
	ds^2 =-e^{\mu}dt^2 + e^{\xi}dr^2 + r^2 (d\theta ^2 + \sin ^2 \theta d\phi^2)~.
\end{equation}
Here $\mu$ and $\xi$ are assumed to be functions of $r$ only. Using the ansatz metric, the Einstein equations take the form \cite{48}
\begin{eqnarray}\label{34}
	e^{-\xi}\Big(\frac{1}{r^2}-\frac{\xi^{'}}{r}\Big)-\frac{1}{r^2}=8\pi \Big(-\rho-\frac{Q^2}{8\pi r^4}\Big)\nonumber\\
	e^{-\xi}\Big(\frac{1}{r^2}+\frac{\mu^{'}}{r}\Big)-\frac{1}{r^2}=8\pi \Big(-\rho-\frac{Q^2}{8\pi r^4}\Big)\nonumber\\
	\frac{e^{-\xi}}{2}\Big(\mu{''}+\frac{\mu{'}^2}{2}+\frac{\mu{'}-\xi^{'}}{r}-\frac{\mu^{'}\xi^{'}}{2}\Big)=8\pi \Big(P+\frac{Q^2}{8\pi r^4}\Big)\nonumber\\
	\frac{e^{-\xi}}{2}\Big(\mu{''}+\frac{\mu{'}^2}{2}+\frac{\mu{'}-\xi^{'}}{r}-\frac{\mu^{'}\xi^{'}}{2}\Big)=8\pi \Big(P+\frac{Q^2}{8\pi r^4}\Big)~.
\end{eqnarray}
Here prime ($'$) and double prime ($''$) denote the first and second derivatives with respect to $r$. Upon rearrangement, the first and third equations respectively can be recast as
\begin{eqnarray}\label{35}
	e^{-\xi}\Big(\frac{1}{r^2}-\frac{\xi^{'}}{r}\Big)-\frac{1}{r^2}+\frac{Q^2}{ r^4}=-8\pi\rho 
 \end{eqnarray}
 \begin{eqnarray}\label{36}
\frac{e^{-\xi}}{2}\Big(\mu{''}+\frac{\mu{'}^2}{2}+\frac{\mu{'}-\xi^{'}}{r}-\frac{\mu'\xi^{'}}{2}\Big)-\frac{Q^2}{ r^4}=8\pi P~. 
\end{eqnarray}
Taking the ratio of eq.\eqref{36} and eq.\eqref{35}, we obtain
\begin{eqnarray}
	\frac{e^{-\xi}}{2}\Big(\mu{''}+\frac{\mu{'}^2}{2}+\frac{\mu{'}-\xi^{'}}{r}-\frac{\mu'\xi^{'}}{2}\Big)-\frac{Q^2}{ r^4}=-\frac{1}{2}\Bigg[e^{-\xi}\Big(\frac{1}{r^2}-\frac{\xi^{'}}{r}\Big)-\frac{1}{r^2}+\frac{Q^2}{ r^4}\Bigg]~.
\end{eqnarray}
Assuming $\mu=-\xi=\ln(1-K)$, where $K\equiv K(r)$, the above equation simplifies to
\begin{equation}
	r^2 K^{''}+3rK^{'} + K + \frac{Q^2}{r^2}=0~.
\end{equation}
The solution of the above equation is 
\begin{equation}
	K(r)=\frac{r_g}{r}-\frac{Q^2}{r^2}-\frac{\chi}{r}ln\Big(\frac{r}{|\chi|}\Big)
\end{equation}
where $r_g$ and $\chi$ are integration constants. To obtain the value of $r_g$, we set $Q=0$ and $\chi=0$. In this limit, we can use the weak field approximation to obtain $r_g=2M$. Therefore the lapse function takes the form
\begin{eqnarray}\label{40}
	f(r)= e^{\mu}=e^{-\xi}=e^{ln(1-K)}=1-K=1-\frac{2M}{r}+\frac{Q^2}{r^2}+\frac{\chi}{r}ln\Big(\frac{r}{|\chi|}\Big)~.
\end{eqnarray}

\noindent In the lapse function, $M$ is the black hole mass, $Q$ is the charge due to the electromagnetic field and $\chi$ is the dark matter parameter related to the energy density $\rho$ of PFDM as $\rho = \frac{\chi}{8\pi r^3}$ \cite{48}. The expression of $\rho$ can be obtained by replacing eq.\eqref{40} in the first equation of eq.\eqref{34}. $\chi$ gives the mass of PFDM enclosed within radius $r$. 

\noindent For our calculational purposes, we rescale all parameters by the black hole mass $M$ which is equivalent to setting $M$ to unity. So in all discussions and results, we omit $M$ but keep the mass $M$ in the plot labels. The \textit{event horizon} of the black hole can be obtained by using the condition 
\begin{equation}\label{41}
   f(r)\Big|_{r=r_+}=0 ~~\Rightarrow ~~ r_+ ^2 -2r_+ + Q^2 + \chi r_+ ln \Big(\frac{r_+}{|\chi|}\Big)=0~.
\end{equation}

\noindent The concerned system is quite unique. The black hole size (event horizon) is dictated by the PFDM parameter $\chi$. Analyzing the lapse function, we found that $r_+$ initially decreases with an increase in $\chi$ reaches a minimum, and then starts to increase. This nature can be observed in the left plot in Fig. \ref{1ab}.

\noindent  The reason for such an observation can be explained by supposing the system to be composed of two masses, one coming from the black hole, $M$, and the other coming from PFDM, $M_0$ \cite{44}. Below and up to $\chi_c$ ($\chi \leq \chi_c$), $M_0$ opposes the effect of the black hole mass $M$, and thereby the event horizon radius $r_+$ decreases up to $\chi_c$. But beyond $\chi_c$ ($\chi > \chi_c$), the total mass of the system (BH + PFDM) is dictated by $M_0$ and hence the event horizon of the black hole-dark matter system increases.  
\begin{figure}[H]
  \centering
  \begin{minipage}[b]{0.4\textwidth}
   {\includegraphics[width=\textwidth]{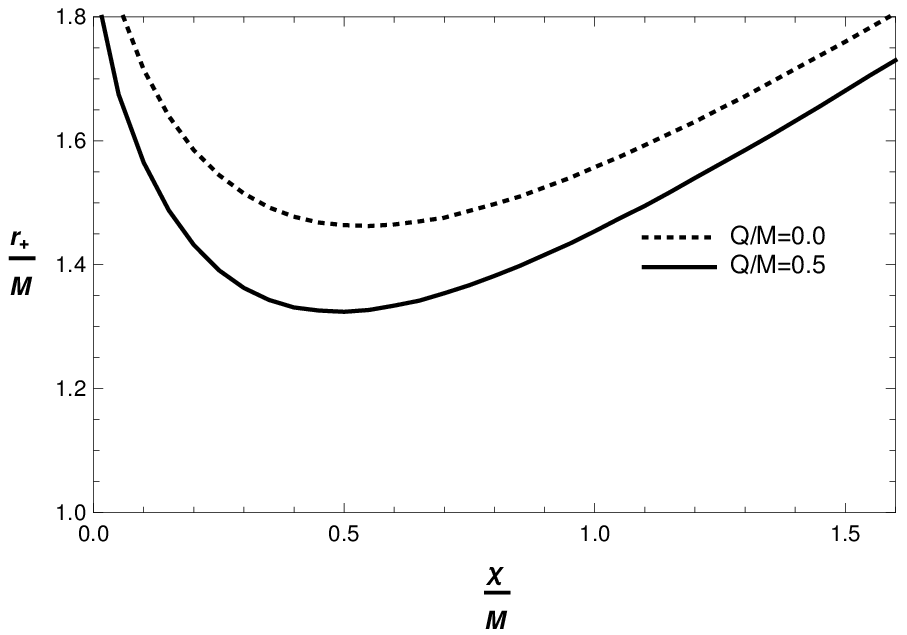}}
    \end{minipage}
  \hspace{1.0cm}
   \begin{minipage}[b]{0.4\textwidth}
    {\includegraphics[width=\textwidth]{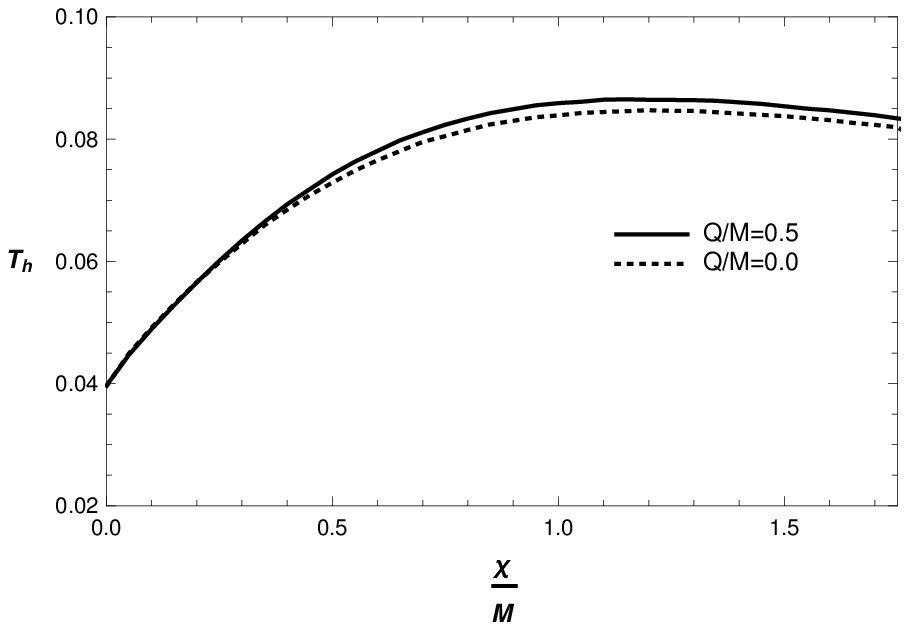}}
     \end{minipage}
  \caption{\footnotesize Plots showing variation of the event horizon ($r_+$) and temperature $T_h$ of the black hole with variation in PFDM parameter $\chi$. The plots are shown for $\frac{Q}{M}=0.0$ (black) and $\frac{Q}{M}=0.5$ (dotted black)~. }
  \label{1ab}
\end{figure}

\noindent The minimum of the event horizon can be obtained using the condition
\begin{equation}\label{42}
    \frac{\partial r_+}{\partial \chi}\Bigg|_{\chi_c}=0~~;~~~\frac{\partial^2 r_+}{\partial \chi ^2}\Bigg|_{\chi_c} > 0
\end{equation}
where $\chi_c$ is the value of $\chi$ corresponding to the minimum of the event horizon radius $(r_+)_c$. Using eq.(s) \eqref{41}, \eqref{42}, we obtain 
\begin{equation}\label{43}
    \chi_{c}=\frac{1}{1+e}\Bigg(1+\sqrt{1-Q^2 \Big(1-\frac{1}{e}\Big)}\Bigg)~;~(r_+)_c=\frac{e}{1+e}\Bigg(1+\sqrt{1-Q^2 \Big(1-\frac{1}{e}\Big)}\Bigg)~.
\end{equation}
In the limit of $Q \to 0$, we get $\chi_{c} = \frac{2}{1+e}$ \cite{23}, \cite{44}. In order to derive eq.\eqref{43}, we differentiate eq.\eqref{41} with respect to $\chi$. Then we use eq.\eqref{42} which gives the relation
\begin{equation}\label{44a}
    (r_+)_c = \chi_c e ~.
\end{equation}
Using the above relation in eq.\eqref{41} and eliminating $(r_+)_c$, we obtain $\chi_c$. Then using $\chi_c$ in eq.\eqref{44a} we get $(r_+)_c$.

\noindent The expression of the temperature of the black hole takes the form
\begin{eqnarray}\label{22}
\nonumber
    T_h &=&\frac{f'(r_+)}{4\pi}\\ \nonumber
    &=&\frac{1}{4\pi}\Bigg[\frac{2+\chi}{r_+ ^2}-\frac{2Q^2}{r_+ ^3}-\frac{\chi}{r_+ ^2}ln\Big(\frac{r_+}{|\chi|}\Big)\Bigg]\\ 
    &=&\frac{1}{4\pi r_+ ^3}\Big[r_+ ^2 +\chi r_+ - Q^2\Big]~.
\end{eqnarray}
Setting $\chi = 0$, we get back the results of the Reissner-Nordstr\"om black hole \cite{16a}
\begin{equation}\label{18}
    T_h=\frac{1}{4\pi}\Bigg[\frac{2}{r_+ ^2}-\frac{2Q^2}{r_+ ^3}\Bigg]~.
\end{equation}

\noindent In Fig.\ref{1ab}, we have shown the variation of the black hole event horizon ($r_+$) (left) and black hole temperature ($T_h$) (right) with increment in the event horizon radius $r_+$. The plots are shown for $Q=0.0$ (black dotted) and $Q=0.5$ (black). 

\noindent The nature of variation of the left plot in Fig.\ref{1ab} has been explained above. Also, we found that the event horizon has a lower value in the presence of a black hole charge $Q$. The reason is, the presence of charge $Q$ reduces the effective mass of the black hole system which gets reflected in the corresponding reduction in the radius of the event horizon ($r_+$).

\noindent The right plot in Fig. \ref{1ab} shows the variation of the black hole temperature ($T_h$) with increase in PFDM parameter $\chi$. We find that $T_h$ initially increases with an increment in $\chi$, reaches a maximum and then starts to decrease. The reason for such an observation is that the black hole temperature gets reduced with an increase in mass. In our system, the effective mass of the black hole system (BH + PFDM) initially decreases with $\chi$, and thus the temperature increases. But after a critical value, the effective mass of the system increases resulting in the decrease of the black hole temperature. Also, we find that in the presence of charge $Q$, the black hole temperature increases. The reason being the effective mass of the black hole system gets reduced due to the presence of charge $Q$ which results in an increment in the black hole temperature.

\subsection{Time-like geodesics}
The radius of the circular time-like geodesics is designated as $r_0$ and the corresponding energy and angular momentum per unit mass as $E_0$ and $L_0$. The lapse function and its derivative takes the form 
\begin{equation}
  f(r)=1-\frac{2}{r}+ \Big(\frac{Q}{r}\Big)^2 + \frac{\chi}{r}ln\Big(\frac{r}{|\chi|}\Big)~~;~~f'(r)=\frac{2+\chi}{r^2}-\frac{2Q^2}{r^3} - \frac{\chi}{r^2}ln\Big(\frac{r}{|\chi|}\Big)~.
\end{equation}
Using the expressions of $f(r)$ and $f'(r)$ in eq.\eqref{6}, we get  $E_0$ and $L_0$ to be
\begin{equation}\label{48}
    E_0 ^2 = \frac{2\Bigg[r_0 ^2 -2r_0 + Q^2 + \chi r_0 ln\Big(\frac{r_0}{|\chi|}\Big)\Bigg]^2}{r_0 ^2 \Bigg[2r_0 ^2 -(\chi+6)r_0 + 4Q^2 + 3\chi r_0ln\Big(\frac{r_0}{|\chi|}\Big)\Bigg]}~~;~~L_0 ^2 = \frac{r_0 ^2 \Bigg[\Big(\chi+2\Big)r_0 - 2Q^2 - \chi r_0 ln\Big(\frac{r_0}{|\chi|}\Big)\Bigg]}{\Bigg[2r_0 ^2 -(\chi + 6)r_0 + 4Q^2 + 3\chi r_0ln\Big(\frac{r_0}{|\chi|}\Big)\Bigg]}~.
\end{equation}
For the energy $E_0$ and angular momentum $L_0$ to be real and finite, we must simultaneously have
\begin{eqnarray}\label{49}
   \Bigg[\Big(\chi+2\Big)r_0 - 2Q^2 - \chi r_0 ln\Big(\frac{r_0}{|\chi|}\Big)\Bigg] \equiv A>0
\end{eqnarray}
\begin{eqnarray}\label{50}
   \Bigg[2r_0 ^2 -(\chi+6)r_0 + 4Q^2 + 3\chi r_0ln\Big(\frac{r_0}{|\chi|}\Big)\Bigg] \equiv B>0~.
\end{eqnarray}
Also we define 
\begin{equation}
  \Bigg[r_0 ^2 -2r_0 + Q^2 + \chi r_0 ln\Big(\frac{r_0}{|\chi|}\Big)\Bigg] \equiv C  
\end{equation}
which gives 
\begin{equation}
    E_0 = \sqrt{\frac{2}{B}}\frac{C}{r_0}~~;~~L_0=r_0 \sqrt{\frac{A}{B}}~~~ \Rightarrow \frac{L_0}{E_0}=\frac{r_0 ^2}{C}\sqrt{\frac{A}{2}}~.
\end{equation}

\noindent Eq.(s) \eqref{49}, \eqref{50} are the conditions for the existence of particles with real and finite values of energy $E_0$ and angular momentum $L_0$. 

\noindent The plot in Fig.\ref{1aba} gives us the region of existence of real and finite $E_0$ and $L_0$ compatible with the possible values of $\chi$ and $r_0$. The compatible range of $\chi$ and $r_0$ are  shown for $Q= 0.0$ (blue), $0.5$ (red), and $1.0$ (magenta). We find that with an increase in charge $Q$ the possible region of existence of circular orbits having finite values of $E$ and $L$ increases. The domain for $Q=0.5$ covers the domain for $Q=0$ along with some extra domain and the compatible region of $\chi$ and $r_0$ for $Q=1.0$ covers the region of both $Q=0.0$ and $Q=0.5$ along with some additional region. We find that with an increase in charge $Q$, the possibility of the existence of orbits closer to the black hole increases. Also, orbits with finite values of $E_0$ and $L_0$ are more probable to be found for lower values of PFDM parameter $\chi$ with increasing charge $Q$. 
\begin{figure}[H]
  \centering
  {\includegraphics[width=0.4\textwidth]{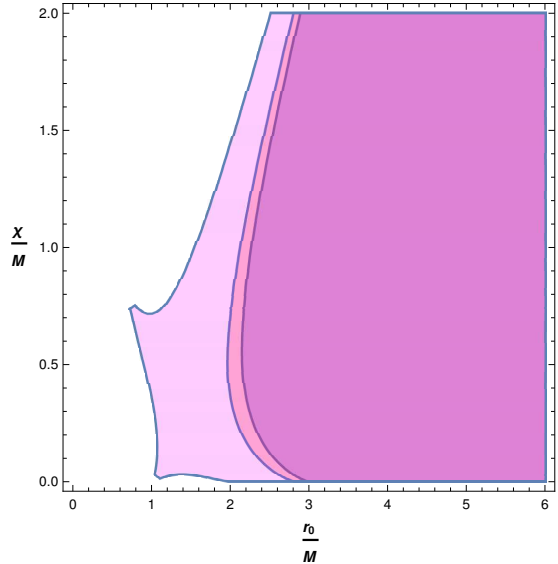}}
  \caption{\footnotesize Parametric plot in ($\frac{\chi}{M}-\frac{r_0}{M}$) plane for the real and finite $E_0$ and $L_0$. The plots are shown for black hole charge $\frac{Q}{M}$=0 (blue), 0.5 (red), and 1.0 (magenta). }
  \label{1aba}
\end{figure}

\noindent Next, we try to find the conditions of \textit{stability/instability}  of the circular geodesics and possible conditions for the observability of those instabilities. In order to analyse the stability we compute the Lyapunov exponent ($\lambda$). As mentioned earlier, there are two Lyapunov exponents in the case of time-like geodesics, one corresponding to co-ordinate time ($t$), $\lambda_{0c}$ and the other to the proper time ($\tau$), $\lambda_{0p}$. 

\noindent In order to compute the co-ordinate time Lyapunov exponent $\lambda_{0c}$, we use the expression in eq.\eqref{L2}. Here we replace the potential $V_r$ from eq.\eqref{7a} with the lapse function $f(r)$ given by eq.\eqref{40}. Also, we use the geodesic equation for $t$ as obtained in eq.\eqref{3}. Since the Lyapunov exponent is evaluated at the radius $r_0$ for the circular orbits, so we replace $r$ by $r_0$ and the corresponding values for $E$ and $L$ by $E_0$ and $L_0$ respectively. Finally, we use the expressions of $E_0$ and $L_0$ as obtained in eq.\eqref{48} to obtain the expression of $\lambda_{0c}$ as
\begin{eqnarray}\label{54a}
    \lambda_{0c}&=&\frac{1}{\sqrt{2r_0 ^6}}\Bigg[r_0 ^3 \Big(\chi ln(\frac{r_0}{|\chi|})
    -2\Big)+ r_0 ^2 \Bigg(2\chi^2 + 8\chi +12 -12\chi ln\left(\frac{r_0}{|\chi|}\right)\nonumber\\
    &-&4\chi^2 (\frac{r_0}{|\chi|})+3\chi^2\Big(ln(\frac{r_0}{|\chi|})\Big)^2\Bigg)  
     + r_{0}\Big(9\chi Q^2ln\left(\frac{r_0}{|\chi|}\right)-18Q^2-8\chi Q^2\Big)+8Q^4\Bigg]^{\frac{1}{2}}\\
\nonumber
&=&\sqrt{\frac{\Delta}{2r_0 ^6}}~.
\end{eqnarray}
In the case of proper time Lyapunov exponent $\lambda_{0p}$ we use the expression in eq.\eqref{L1}. Then we replace the potential $V_r$ using eq.\eqref{7a}. We use the lapse function $f(r)$ as given in eq.\eqref{40}. Since the expression for $\lambda_{0p}$ is evaluated at the circular radius $r_0$, so we replace the $E$ and $L$ by $E_0$ and $L_0$ respectively and put their values as calculated in eq.\eqref{48} to obtain

\begin{eqnarray}\label{55}
    \lambda_{0p}&=& \frac{1}{\sqrt{r_0 ^4  B}}\Bigg[r_0 ^3 \Big(\chi ln(\frac{r_0}{|\chi|})-2\Big)+ r_0 ^2 \Bigg(2\chi^2 + 8\chi +12 -12\chi ln\left(\frac{r_0}{|\chi|}\right)\nonumber\\
    &-&4\chi^2 \left(\frac{r_0}{|\chi|}\right)+3\chi^2\Big(ln\left(\frac{r_0}{|\chi|}\right)\Big)^2\Bigg)  
     + r_{0}\Big(9 \chi Q^2ln\left(\frac{r_0}{|\chi|}\right)-18Q^2-8 \chi Q^2\Big)+8Q^4\Bigg]^{\frac{1}{2}} \nonumber\\
     &=&\sqrt{\frac{\Delta}{r_0 ^4 B}}~.
\end{eqnarray}
\noindent Previously we mentioned that $B>0$ is a required condition for the existence of particles with finite values of $E$ and $L$. Together with those conditions, we also require $\Delta<0$, which results in imaginary $\lambda$  and corresponds to the stability of the geodesic \cite{20a}. On the other hand for $\Delta>0$, we have a real $\lambda$ which corresponds to unstable geodesics. And the condition of $\Delta=0$ gives the marginal circular geodesic or the innermost stable circular geodesic.
\begin{figure}[H]
    \centering
      {\includegraphics[width=0.4\textwidth]{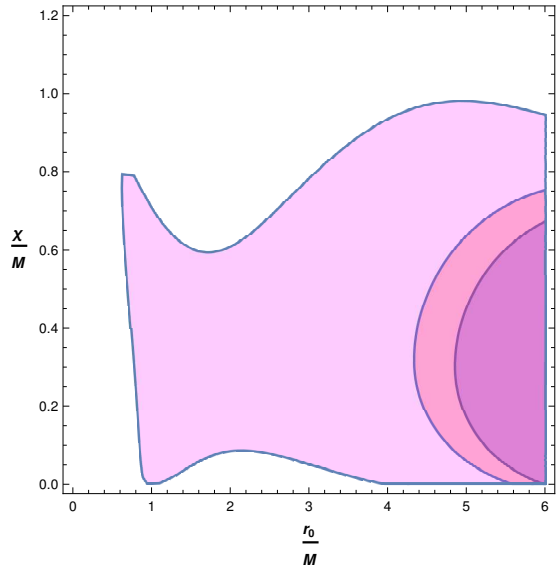}}
      \caption{\footnotesize Parametric plot in ($\frac{\chi}{M}-\frac{r_0}{M}$) plane for the existence of stable orbit. The plots are shown for black hole charge $\frac{Q}{M}$=0 (blue), 0.5 (red), and 1.0 (magenta).}
    \label{aaa1}
\end{figure}
\noindent The plot in Fig.\ref{aaa1} shows the region of existence of stable orbits of radius $r_0$ compatible with the values of $\chi$. The region increases with an increase in $Q$, thereby increasing the possible compatible range of $\chi$ and $r_0$. The plots are shown for $Q=0.0$ (blue), $0.5$ (red), and $1.0$ (magenta). The region for $Q=0.5$ (red) covers the region accessible in case of $Q=0.0$ (blue) and the domain in case of $Q=1.0$ (magenta) covers the domain for both $Q=0.0$ (blue) and $Q=0.5$ (red). We observe an increase in the range of accessible stable orbits of radius $r_0$. The stable orbits get closer to the black hole with an increase in charge $Q$. Due to the increase in charge $Q$, the event horizon and thereby the size of the black hole decreases. Hence the possibility of the existence of orbits closer to the black hole increases. These effectively led to the existence of stable orbits in the region closer to the black hole.

\noindent In order to analyse the observability of gravitational signals, we need to determine the critical exponent $\gamma$. As defined in eq.\eqref{032} $\gamma$ depends on the angular velocity $\Omega$ and Lyapunov exponent $\lambda$. We first calculate the angular velocity $\Omega$ which takes the form
\begin{equation}
        \Omega_0 = \frac{d \phi}{dt}\Big|_{r=r_0}=\sqrt{\frac{f'(r_0)}{2r_0}}=\sqrt{\frac{\Big(\chi+2\Big)r_0 - 2Q^2 - \chi r_0 ln\Big(\frac{r_0}{|\chi|}\Big)}{2r_0 ^4}}=\sqrt{\frac{A}{2r_0 ^4}}~.
\end{equation}
Again the critical exponent $\gamma$ is related to two different time periods whose values give us an idea of the observability of gravitational wave signals \cite{49b}. The critical exponent $\gamma_{0c}$ (for co-ordinate time) and $\gamma_{0p}$ (for proper time)  takes the form
\begin{equation}
    \gamma_{0c}=\frac{T_{\lambda_{0c}}}{T_{\Omega}}=\frac{1}{2\pi}\frac{\Omega_0}{\lambda_{0c}}=\frac{1}{2\pi}\sqrt{\frac{r_0 ^2 A}{\Delta}}~~;~~ \gamma_{0p}=\frac{T_{\lambda_{0p}}}{T_{\Omega}}=\frac{1}{2\pi}\frac{\Omega_0}{\lambda_{0p}}=\frac{1}{2\pi}\sqrt{\frac{AB}{2\Delta}}~.
\end{equation}
From all our previous results we have  $A, B, C >0$. Additionally, $\Delta > 0$  results in positive values of the Lyapunov exponents and thereby the existence of unstable circular geodesics. Also in order to have observational relevance, $\gamma$ must be less than 1, leading to $T_{\lambda} < T_{\Omega}$ \cite{70}. We can therefore find the parameter space of ($r_0, Q, \chi$) which corresponds to such a condition.
\begin{figure}[H]
  \centering
  \begin{minipage}[b]{0.4\textwidth}
   {\includegraphics[width=\textwidth]{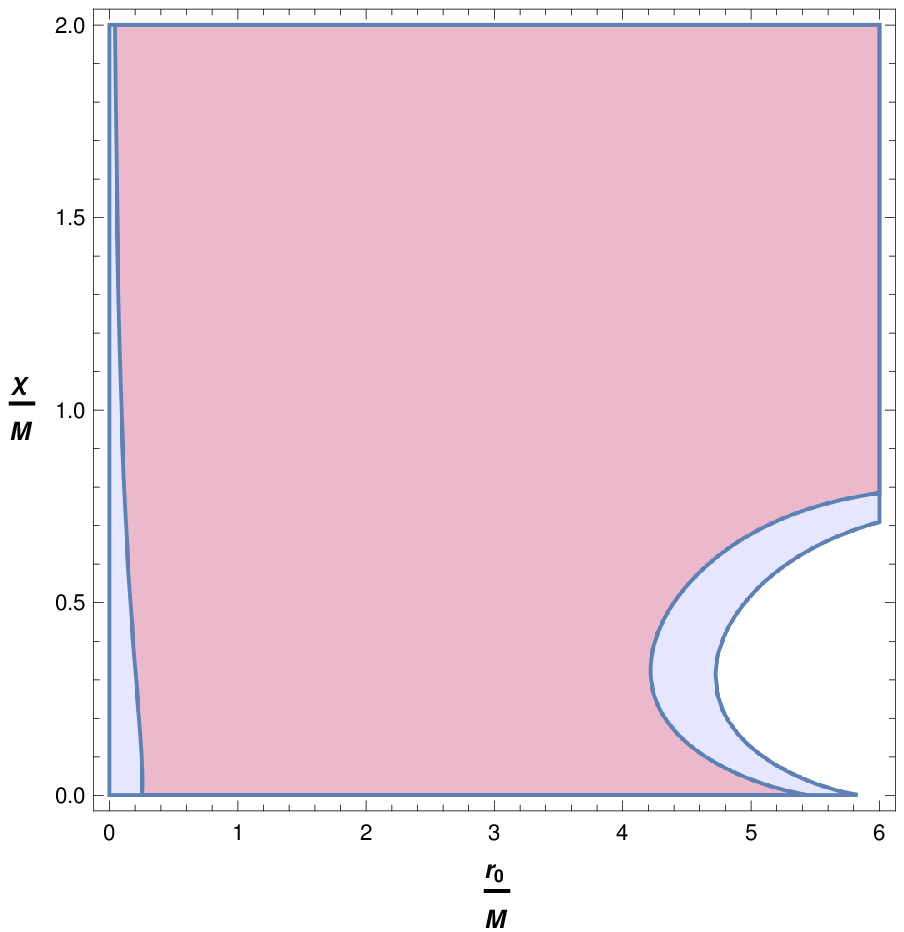}}
    \end{minipage}
  \hspace{1.0cm}
   \begin{minipage}[b]{0.4\textwidth}
    {\includegraphics[width=\textwidth]{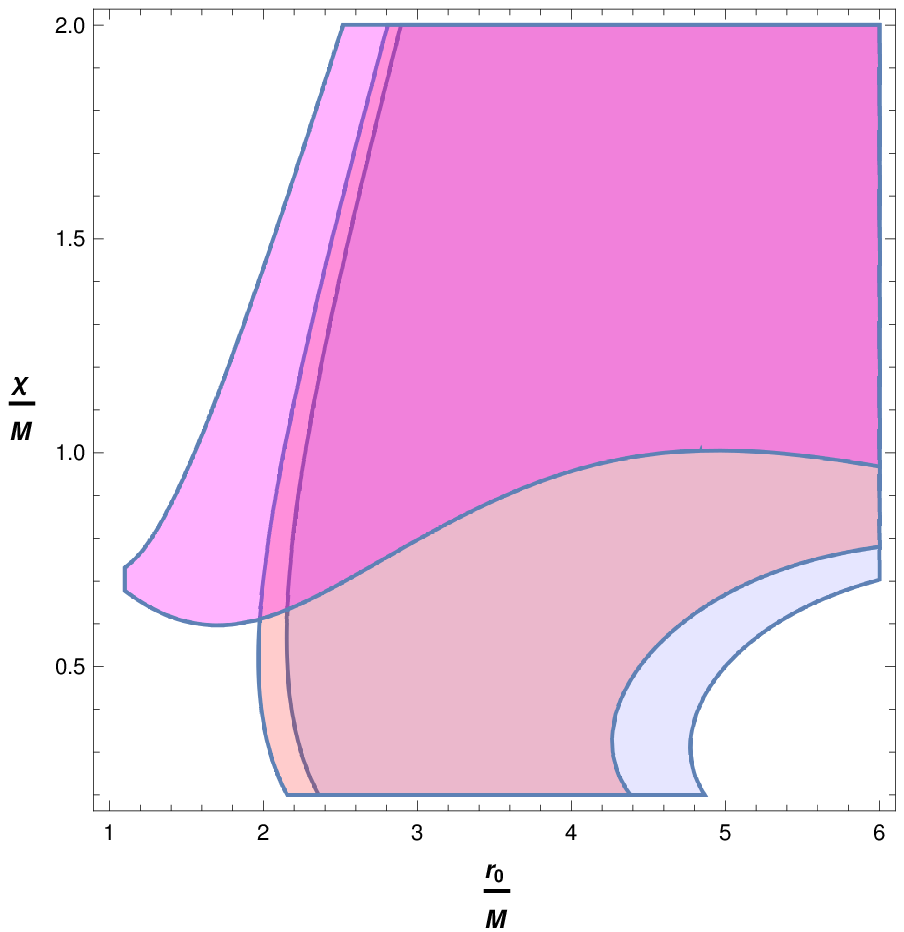}}
     \end{minipage}
  \caption{\footnotesize Parametric plot in ($\frac{\chi}{M}-\frac{r_0}{M}$) plane for possibility of detection of gravitational waves~. The coordinate time critical exponent is on the left whereas the proper time exponent is on the right. The plots are shown for black hole charge $\frac{Q}{M}$=0 (blue), 0.5 (red), and 1.0 (magenta). }
  \label{1bbb}
\end{figure}

\noindent The plots in Fig.\ref{1bbb} show the parameter space corresponding to the detection of gravitational waves. The plots shown, are for charge $Q= 0.0$ (blue), $0.5$ (red), and $1.0$ (magenta). The plots are for coordinate time (left) and proper time (right). We find that the region and thereby the possibility of detection decreases with an increase in black hole charge $Q$. We find that in the case of coordinate time, the domain for $Q=0.5$ (red) almost covers the domain for $Q=0.0$ (blue) with some regions left out. On the other hand in the case of proper time, we find that $Q=0.5$ (red)  almost covers the region covering the entire domain of $Q=0.0$ (blue) along with some extra region for smaller $r_0$ and leaving out some regions of higher values of $r_0$. Also, we find that for $Q=1.0$ (magenta) the region covered is a portion of the domain covered for $Q=0.0$ (blue), $Q=0.5$ (red) along with some additional regions and some left-out regions as well.  So, effectively we find that with an increase in charge $Q$, the smaller values of $r_0$ become more accessible. Also, we find that the detectability of signals and thereby the parameter space of $\chi - r_0$ is larger in the case of coordinate time than that in proper time.

\noindent The coordinate time is more relevant, since it is the time measured by an observer, whereas the proper time is the time as measured in the particle's frame. So, in the case of a coordinate time critical exponent, we find that the entire range of $\chi$ is accessible, that is the detection of gravitational signals does not put any constraint on $\chi$. On the other hand, we find that unstable orbits exist up to a certain region. So, the possibility of detecting gravitational signals constrains $r_0$. Also, we find that the possible values of $r_0$ decrease with $Q$. The reason being with an increase in black hole charge $Q$, the event horizon and thereby the size of the black hole decreases. Thus, the possibility of the existence of all kinds of orbits closer to the black hole increases. With the increase in charge $Q$, the region of existence of unstable orbits $r_0$ decreases.

\subsection{Null geodesics}
Here we are interested in analyzing the geodesics of null particles (photons) for which $\beta=0$. The radial geodesic equation is of the form
\begin{equation}\label{5a}
    \dot{r}^2 = E^2 - f(r)\frac{L^2}{r^2} =E^2 - \Bigg(1-\frac{2M}{r}+ \Big(\frac{Q}{r}\Big)^2 + \frac{\chi}{r}ln\Big(\frac{r}{|\chi|}\Big)\Bigg)\frac{L^2}{r^2} =\widetilde{V}_{r} ~.
\end{equation}
Using the condition for null circular geodesics, that is
\begin{equation}\label{r}
    \dot{r}^2\Bigg|_{r=r_p}=(\dot{r}^2)'\Bigg |_{r=r_p}=0
\end{equation}
we get two constraint equations. 

\noindent The second condition in eq.\eqref{r} gives
\begin{equation}\label{31}
    r_p f'(r_p)-2f(r_p)=0~~ \Rightarrow ~~2r_p ^2 -(\chi+6)r_p + 4Q^2 + 3\chi r_pln\Big(\frac{r_p}{|\chi|}\Big)=0~.
\end{equation}
\noindent This equation can be solved to determine the photon sphere radius $r_p$. But since the above equation is a transcendental equation, it cannot be solved analytically and has to be solved numerically to get $r_p = r_p (Q, \chi)$. The first condition in eq.(\eqref{r}) results in a constraint on $L$ and $E$ in terms of $r_p$ as
\begin{equation}\label{32}
    \frac{L_p}{E_p}= \sqrt{\frac{f(r_p)}{r_p ^2}} = \sqrt{\frac{r_p ^2 -2r_p + Q^2 + \chi r_p ln\Big(\frac{r_p}{|\chi|}\Big)}{r_p ^4}}=\sqrt{\frac{r_p ^2 + \chi r_p -Q^2}{3 r_p ^4}}~.
\end{equation}
The last expression in eq.\eqref{32} is obtained using eq.\eqref{31} and replacing the logarithmic term. Also, $E_p$ and $L_p$ are energy and angular momentum of null particles moving in circular orbits of radius $r_p$.

\noindent In Fig(\ref{g}), we have shown the variation of photon orbit radius $r_p$ with $\chi$. The plots are shown for $Q =0.0$ (black dashed) and $0.5$ (black). We find that for fixed values of charge $Q$, the photon sphere radius $r_p$ initially decreases with $\chi$ reaches a minimum and then again starts to increase. Also, we find that for a fixed value of $\chi$, $r_p$ decreases with an increase in charge $Q$.

\begin{figure}[H]
\centering
\includegraphics[width=0.4\textwidth]{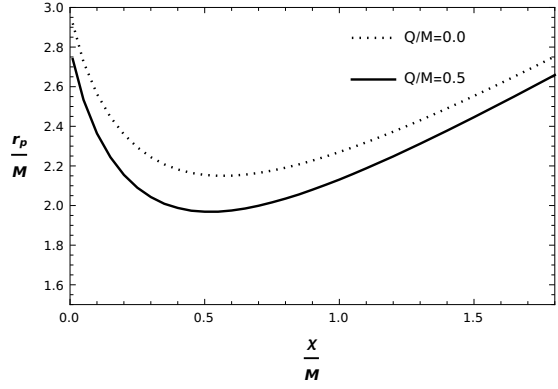}
\caption{\footnotesize Variation of photon orbit radius $\frac{r_p}{M}$ with change in $PFDM$ parameter $\frac{\chi}{M}$. The plots are shown for $\frac{Q}{M} =0$(black) and $\frac{Q}{M}=0.5$ (dotted black).}
\label{g}
\end{figure}

\noindent The reason for the above observation can be understood from the location and process of formation of the photon sphere.  The photon sphere is the sphere composed of unstable photons from which we receive light and is the closest anything can reach a black hole and escape it. The size of the photon sphere is dictated by the size of the event horizon of the black hole which for Schwarzschild black hole is $r_p = 3M = \frac{3}{2}r_+ = 1.5r_+$ which can be obtained from eq.\eqref{31} by setting $\chi = Q=0$. So we find that $r_p$ indirectly gives the size of $r_+$ (the event horizon)  and thereby is expected to have a similar dependence on $\chi$ as that of $r_+$ (discussed earlier) which is reflected in the above Figure. 

\noindent The angular velocity $\Omega_p$ for null circular geodesics takes the form
\begin{equation}\label{33}
    \Omega_p = \frac{d\phi}{dt}\Bigg|_{r=r_p}=\sqrt{\frac{f(r_p)}{r_p ^2}} = \frac{E_p}{L_p}=\sqrt{\frac{r_p ^2 + \chi r_p -Q^2}{3 r_p ^4}}~.
\end{equation}
The above expression can be obtained using the geodesics of $\phi$ and $t$ (eq.\eqref{3}) and then using eq.\eqref{32}. The expression $\frac{L_p}{E_p}=D$ is defined as the impact parameter which is related to the angular velocity of photons moving in circular null geodesics as shown in eq.\eqref{33}.

\noindent In the case of null geodesics, only the co-ordinate time Lyapunov exponent $\lambda_{p}$ can be defined  which takes the form
\begin{eqnarray}\label{65}
\nonumber
    \lambda_{p} &=& \sqrt{\frac{\widetilde{V}''_{r}}{2\dot{t}^2}}\Bigg|_{r=r_p}\\ \nonumber
    &=&\frac{1}{\sqrt{4r_p ^6}}\Bigg[\Bigg((\chi +2)r_p -2Q^2 -\chi r_pln\Big(\frac{r_p}{|\chi|}\Big)\Bigg)\Bigg((6+4\chi)r_p -8Q^2 -3 \chi r_pln\Big(\frac{r_p}{|\chi|}\Big)\Bigg)\Bigg]^{\frac{1}{2}}\\
    &=&\sqrt{\frac{PR}{4r_p ^6}}
\end{eqnarray}
with $P$ and $R$ taking the form
\begin{equation}
    P=(\chi +2)r_p -2Q^2 -\chi r_pln\Big(\frac{r_p}{|\chi|}\Big)~~;~~R=(6+4\chi)r_p -8Q^2 -3 \chi r_pln\Big(\frac{r_p}{|\chi|}\Big)~.
\end{equation}
The above expression for $\lambda_p$ can be obtained using the potential $\widetilde{V}_r$ as in eq.\eqref{5a} and $t$ geodesic (eq.\eqref{3}). Then we use eq. (s)\eqref{31}, \eqref{32} to obtain eq.\eqref{65}. 

\noindent For stable geodesics, we need $ \lambda_{p}$ to be imaginary. Hence we must have either $P> 0$ and $R<0$ or $P< 0$ and $R>0$. Again for unstable geodesics, we must have real $ \lambda_{p}$ which corresponds to $P> 0$ and $R>0$ or $P< 0$ and $R <0$. 

\noindent The critical exponent $\gamma$ takes the form
\begin{equation}
    \gamma=\frac{1}{2\pi}\frac{\Omega_p}{\lambda_{p}}=\frac{1}{2\pi}\sqrt{\frac{4r_p ^2 \Big(r_p ^2 + \chi r_p -Q^2\Big)}{3PR}}
\end{equation}
which can be obtained by using the expressions for $\Omega_p$ and $\lambda_p$ from eq.(s)\eqref{33}, \eqref{65} respectively. Here the suffix p does not correspond to proper time but corresponds to photon sphere radius $r_p$ at which all quantities are evaluated. The condition $\gamma < 1$ leads to the possibility of observation of gravitational signals \cite{70}.

\begin{figure}[H]
\begin{center}
\includegraphics[width=0.4\textwidth]{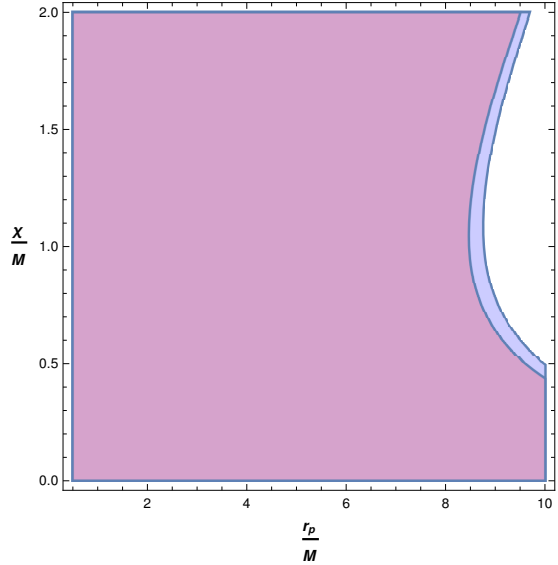}
\end{center} 
{\vspace{-0.4cm}
\caption{\footnotesize Parametric plot in ($\frac{\chi}{M}-\frac{r_p}{M}$) plane for possible detection of gravitational waves. The plots are shown for black hole charge $\frac{Q}{M}$=0 (blue) and 0.5 (red).}
\label{009}}
\end{figure}

\noindent In the above Figure, we have shown the parametric plot of $\chi$ and $r_p$  compatible with the detection of gravitational waves. We observe that the presence of charge $Q$ affects the compatible range of $\chi$ and $r_p$. The plots are shown for $Q= 0$ (blue) and $Q=0.5$ (red). We find that the possible domain for $Q=0.5$ (red) leaves out a small portion of the domain available for $Q=0.0$ (blue). With the increase in black hole charge $Q$, the possible values of $r_p$ decrease. This is because the black hole charge reduces the size of the event horizon $r_+$ which in effect reduces the possible values of unstable null geodesics. Also, we observe that with an increase in $\chi$, the range of possible values of $r_p$ decreases reaching a minimum, and then again starts to increase. This can be explained by the behaviour of $r_+$ which is reflected in the nature of $r_p$.

\section{Perturbation by a scalar field on the black hole background}\label{sec3}
In this section, we are interested in studying the perturbation of a black hole spacetime due to the presence of a massless scalar field $\Phi$. The equation of motion of the perturbing field $\Phi$ in the black hole background takes the form
\begin{equation}\label{37}
   \boxed{\nabla_{\mu} \nabla^{\mu} \Phi =  \frac{1}{\sqrt{-g}}\partial_{\mu} \Bigg(\sqrt{-g}g^{\mu \nu}\partial_{\nu}\Bigg)\Phi=0~.}
\end{equation}
Since we are interested in perturbation around a spherically symmetric metric so the ansatz solution $\Phi$ can be expressed in terms of spherical harmonics $Y_{lm}(\theta, \phi)$ \cite{71}. The solution can be assumed to have the form \cite{16a}, \cite{23}
\begin{eqnarray}
   \Phi = \sum_{l,m} e^{-i\omega t} Y_{lm} (\theta, \phi) \frac{R(r)}{r} ~.
\end{eqnarray}
Replacing the above ansatz in eq.\eqref{37} and using the equation of the \textit{associated Legendre polynomial} \cite{71}, we get a radial equation of the form 
\begin{equation}
    \frac{1}{r^2}\frac{d}{dr}\Bigg[r^2 f(r) \frac{d}{dr}\Bigg(\frac{R(r)}{r}\Bigg)\Bigg] + \Bigg(\frac{\omega^2}{f(r)}-\frac{l(l+1)}{r^2} \Bigg)\frac{R(r)}{r}=0~.
\end{equation}
Using tortoise co-ordinate $r_{*}$ which is related to $r$ as $\frac{dr}{dr_*} = f(r)$, we get the radial equation as 
\begin{equation}\label{1a}
    \frac{d^2 R(r_*)}{dr_{*} ^2} + \Big(\omega^2 - V(r_{*})\Big)R(r_*)=0
\end{equation}
with $V(r_{*})$ giving the potential as
\begin{equation}
    V(r_*)=f(r)\Bigg(\frac{l(l+1)}{r^2} + \frac{f'(r)}{r}\Bigg)~~;~~\frac{dr}{dr_{*}}=f(r) \Rightarrow r_* = \int \frac{dr}{f(r)}~.
\end{equation}
The above equation can be rewritten in the form
\begin{equation}
     \frac{d^2 \psi(x)}{dx^2} + Q(x)\psi(x)=0 
\end{equation}
with $r_* \to x$, $\Big(\omega^2 - V(r_{*})\Big) \to Q(x)$ and $R(r_*) \to \psi (x)$. The above equation is Schr\"odinger-like and can be easily solved for $Q(x)=$constant. But in this case, $Q(x)$ being a function of $x$, the equation can be solved for some specific forms of $Q(x)$ (thereby of $V(r_*)$). The formulation has been detailed in Schutz and Will \cite{72}. They used the semi-analytic WKB approximation to solve the problem.

\noindent In general, the WKB method is valid in the case of slowly varying potentials \cite{73}. So, our first criterion will be that the potential or rather $Q(x)$ must be nearly constant. Again, in problems using $WKB$ approximation, we have an incident, reflected, and transition amplitudes with the incident and reflected ones being comparable. 

\noindent But in the case of black hole perturbations, there are no such incident amplitudes, and thus the reflected and transition amplitudes are comparable. So in such cases, the WKB approximation can be made applicable only if  
\begin{equation}
    Q(x)\Bigg|_{x=x_0}= \frac{dQ(x)}{dx}\Bigg|_{x=x_0}=0~.
\end{equation}
The above condition is similar to the condition of circular geodesics. But the turning points will be too close to be applicable for the $WKB$ approximation. The only possibility is if the matching of the solutions across the two boundaries is done simultaneously. Also, the potential is assumed to be parabolic in nature\footnote{In general, most potentials are parabolic close to the maxima.}. The solutions are matched across the different regions by using Taylor expansion of the potential \cite{73}. 

\noindent We expand $Q(x)$ about the maxima ($x_0$) and find a solution in the region between the two turning points. Since the solution is continuous, we can find the approximate solution in the two regions which are constant at far-off regions, since $Q(x) \to \omega$(= constant) at far regions. This results in a relation termed as \textit{quasi-normal mode condition} or QNM condition in literature since it leads to the existence of quasi-normal modes. The condition takes the form \cite{72}
\begin{equation}
    \frac{Q(x_0)}{\sqrt{2Q^{''} (x_0)}}=-i\Big(n+\frac{1}{2}\Big)~.
\end{equation}
A short derivation of this result is presented in the Appendix.

\noindent Using the above condition in eq.\eqref{1a}, we get 
\begin{equation}\label{44}
    \frac{\omega^2 - V(\tilde{r}_{0})}{\sqrt{-2V^{''} (\tilde{r}_0)}}=-i\Big(n+\frac{1}{2}\Big)~~ \Rightarrow ~~ \omega^2=V(\tilde{r}_{0})-i\Big(n+\frac{1}{2}\Big)\sqrt{-2V^{''} (\tilde{r}_0)}=A-iB
\end{equation}
where $\tilde{r}_0$ corresponds to the extrema of the potential with $A = V(\tilde{r}_{0})$ and $B=\Big(n+\frac{1}{2}\Big)\sqrt{-2V^{''} (\tilde{r}_0)}$. The \textit{exact} expression of the frequency takes the form
\begin{equation}\label{45}
    \omega= \omega_R - i\omega_I~~;~~ \omega_R=\sqrt{\frac{A+\sqrt{A^2 + B^2}}{2}}~;~ \omega_I=\frac{B}{\sqrt{2\Big(A+\sqrt{A^2 + B^2}\Big)}}~.
\end{equation}
The condition for determining the extremum point $\tilde{r}_0$ of the potential takes the form
\begin{eqnarray}\label{45a}
    \frac{l(l+1)}{\tilde{r}_0 ^3}\Big(\tilde{r}_0 f'(\tilde{r}_0)-2f(\tilde{r}_0)\Big) + \Big(\frac{(f'(\tilde{r}_0))^2} {\tilde{r}_0}+\frac{f''(\tilde{r}_0)f(\tilde{r}_0)}{\tilde{r}_0}-\frac{f'(\tilde{r}_0)f(\tilde{r}_0)}{\tilde{r}_0 ^2}\Big)=0~.
\end{eqnarray}
The above eq.\eqref{45a} in case of the Schwarzschild black hole takes the form
\begin{equation}
   \tilde{r}_{0S} ^2 - \Big(3-\frac{4}{l(l+1)}\Big)\tilde{r}_{0S} - \frac{10}{l(l+1)} = 0~. 
\end{equation}
$ \tilde{r}_{0S} $ is the point of maxima of the potential in the Schwarzschild background. Since the above equation is quadratic in nature, so one can obtain an analytical solution that has the form
\begin{equation}
    \tilde{r}_{0S} =\frac{1}{2}\Bigg(\Big(3-\frac{4}{l(l+1)}\Big) + \sqrt{\Big(3-\frac{4}{l(l+1)}\Big)^2 + \frac{40}{l(l+1)}} \Bigg)~.
\end{equation}
The above solution in the limit of large $l$, that is $l \to \infty$, results in $\tilde{r}_{0S} = 3$.
For Reissner-Nordstr\"om black hole, eq.\eqref{45a} takes the form
\begin{equation}
    \tilde{r}_{0RN}^4 - \Big(3- \frac{4}{l(l+1)}\Big)\tilde{r}_{0RN}^3 + \Big(2Q^2 -\frac{10(1 + Q^2)}{l(l+1)}\Big)\tilde{r}_{0RN}^2 + \frac{18Q^2}{l(l+1)} \tilde{r}_{0RN} - \frac{7Q^4}{l(l+1)} =0
\end{equation}
 $\tilde{r}_{0RN}$ is the point of maxima of the potential in the Reissner-Norstr\" om background. The above equation is quartic and needs to be solved numerically. In the limit of large $l$, that is $l \to \infty$, we have the equation
\begin{equation}
    \tilde{r}_{0RN}^2 \Big(\tilde{r}_{0RN}^2 - 3\tilde{r}_{0RN} + 2Q^2\Big)= 0 ~.
\end{equation}
The solution of the above equation gives
\begin{equation}
    \tilde{r}_{0RN}=\frac{3}{2}\Bigg(1 + \sqrt{1-\frac{8}{9}Q^2}\Bigg)~.
\end{equation}
The equation for determining $\tilde{r}_0$ in case of charged black hole immersed in PFDM is complicated (not shown) and can be solved numerically for general $l$ as well as for $l>>>1$. We would also like to point out that the above results are shown for the first-order WKB approximation which is followed all throughout our analysis for simplicity. The higher order WKB approximations and the corresponding expressions for quasinormal frequency $\omega$ are obtained in \cite{74}, \cite{75}, \cite{76}.

\subsection{Eikonal approximation}
The eikonal approximation is used to approximate the solutions of certain partial differential equations
(PDEs), particularly those which arise in the study of wave propagation problems. In the eikonal approximation, the main idea is to neglect the wave nature of a propagating
wave and consider only its geometrical properties. This approximation assumes that the wavelength of
the wave is much smaller compared to the characteristic length scale of the system under consideration.
This approximation simplifies calculations and provides insight into the propagation of waves. In optics,
the eikonal approximation is often used to study the behavior of light rays in refractive media or near
optical interfaces \cite{e1}. The eikonal approximation can also be used to determine the path of light rays \cite{e2}, \cite{e3}, and calculate the bending of light at boundaries \cite{e2}, \cite{e4}, to name a few. This approximation is also very useful in the study of scattering problems in quantum mechanics \cite{73}, \cite{e5}. \\
\noindent The eikonal approximation is used to study the quasinormal modes of a black hole \cite{50}. These modes are
important for understanding the behaviour of black holes and testing various theories of gravity. In the eikonal
approximation, the black hole is treated as a geometric object with a specific shape and mass distribution.
The quasinormal modes can then be calculated by solving the wave equation for small perturbations around
the black hole geometry. The eikonal approximation assumes that the wavelength of the perturbation is much smaller than the black hole
horizon radius. The eikonal approximation leads to an approximate formula for the quasinormal frequencies
of the black hole, which depends on the properties of the black hole, such as its mass, spin, and charge.
It provides an estimate of the quasinormal decay rate, which describe how quickly the oscillations of the black hole decay over time. The decay rate is related to the imaginary part
of the quasinormal frequency. It is important to note that the eikonal approximation may break down if
the perturbations are too large, or if the black hole is spinning rapidly. However, it remains a useful tool
for studying the quasinormal modes of black holes and understanding the behaviour of gravity in the strong
gravitational fields near a black hole.

\noindent In order to incorporate the eikonal approximation in our analysis, we impose the condition $l>>1$ \cite{50}. This results into an approximated potential $V(r_{*})$ of the form
\begin{equation}\label{51}
    V(r_{*})=f(r)\frac{l(l+1)}{r^2}~. 
\end{equation}
The condition for the maxima of the potential ($V'(r_*)|_{r=r_0}=0$) gives
\begin{equation}\label{52a}
    \frac{l(l+1)}{\tilde{r}_0 ^3}\Big(\tilde{r}_0 f'(\tilde{r}_0)-2f(\tilde{r}_0)\Big)=0~.
\end{equation}

\noindent We find that both the  eq.(s) \eqref{7}, \eqref{52a} correspond to the same condition for any arbitrary lapse function $f(r)$. \textit{Thus in the eikonal limit, the extremum point ($\tilde{r}_0$) of the function $\Big(\omega^2 - V(r_*)\Big)$ corresponds to unstable circular null geodesics ($r_p$).} Hence, we can get an idea of the quasi-normal frequencies from the knowledge of unstable circular null geodesics.

\noindent The eikonal approximation results in the quasi-normal frequencies 
\begin{equation}\label{52}
    \omega^2 = V(r_p)-i(n+\frac{1}{2})\sqrt{-2V''(r_p)} = l(l+1) \Omega_p ^2 -i2\Big(n+\frac{1}{2}\Big)\sqrt{l(l+1)}\lambda\Omega_p = C-iD
\end{equation}
where the quasinormal frequencies can be represented as $\omega = \Tilde{\omega}_R - i\Tilde{\omega}_I $ and which in terms of $C$ and $D$ take the form
\begin{equation}
   \widetilde{\omega}_R=\sqrt{\frac{C+\sqrt{C^2 + D^2}}{2}}~;~ \widetilde{\omega}_I=\frac{D}{\sqrt{2\Big(C+\sqrt{C^2 + D^2}\Big)}}~.
\end{equation}
The final form of equation eq.\eqref{52} can be obtained by using eq.(s) \eqref{32}, \eqref{33}, \eqref{65}.

\noindent We wish to calculate the values of $\omega_R$ and $\omega_I$ using the eikonal approximation that is using $r_p$ and compare with the same evaluated using $\tilde{r}_0$. As mentioned above, we obtain our results using the first-order $WKB$ approximation. The higher-order corrections would be performed in our future works. Previous works have been done using PFDM using the sixth order \cite{23} but no such comparisons are shown. Besides in our work we carry out the analysis using the charged black hole.

\noindent In the eikonal limit, a further approximation is possible in eq.\eqref{52}. We can assume that $l >> n$ and since $l$ is large, thus $l(l+1) \sim l^2$. Hence, eq. \eqref{52} simplifies to \cite{50}
\begin{equation}\label{54}
    \omega = l \Omega_p - i(n+\frac{1}{2})|\lambda|~.
\end{equation}
We find in the next section that the black hole shadow radius $R_s$ takes the form $R_s = \frac{r_p}{\sqrt{f(r_p)}}= \Omega_p ^{-1}$. Thus the quasinormal frequency $\omega$ can be written as \cite{23}
\begin{equation}
    \omega = \frac{l}{R_s} - i(n+\frac{1}{2})|\lambda|~.
\end{equation}
Also, the Lyapunov exponent $\lambda$ can also be related to the shadow radius $R_s$ along with some other parameters as shown in \cite{23}. Thus the shadow radius and QNM frequency are interrelated and one can be obtained with knowledge of the other. Besides, since both are measurable quantities, hence the system of interest can be cross-checked using both results.

\begin{figure}[H]
  \centering
  \begin{minipage}[b]{0.4\textwidth}
   {\includegraphics[width=\textwidth]{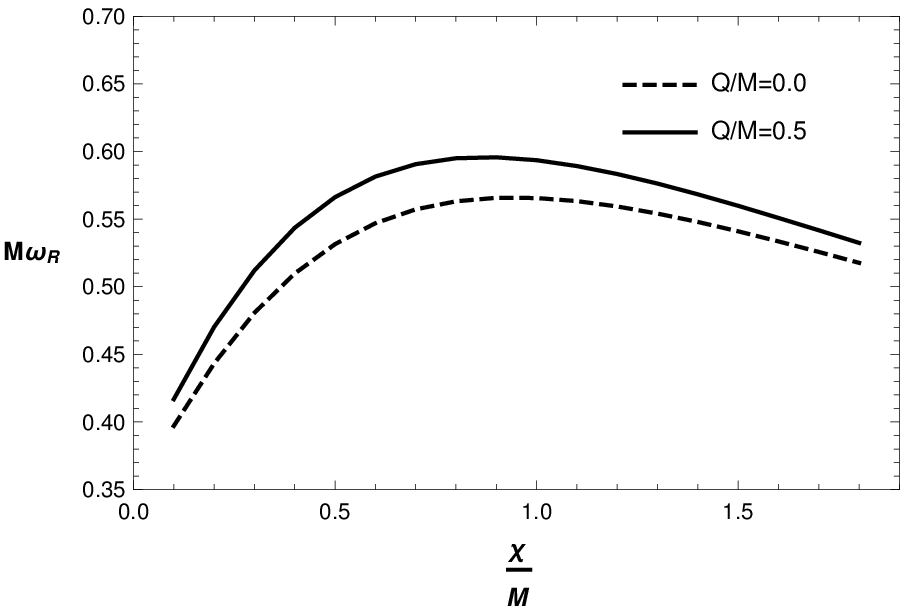}}
    \end{minipage}
  \hspace{1.0cm}
   \begin{minipage}[b]{0.4\textwidth}
    {\includegraphics[width=\textwidth]{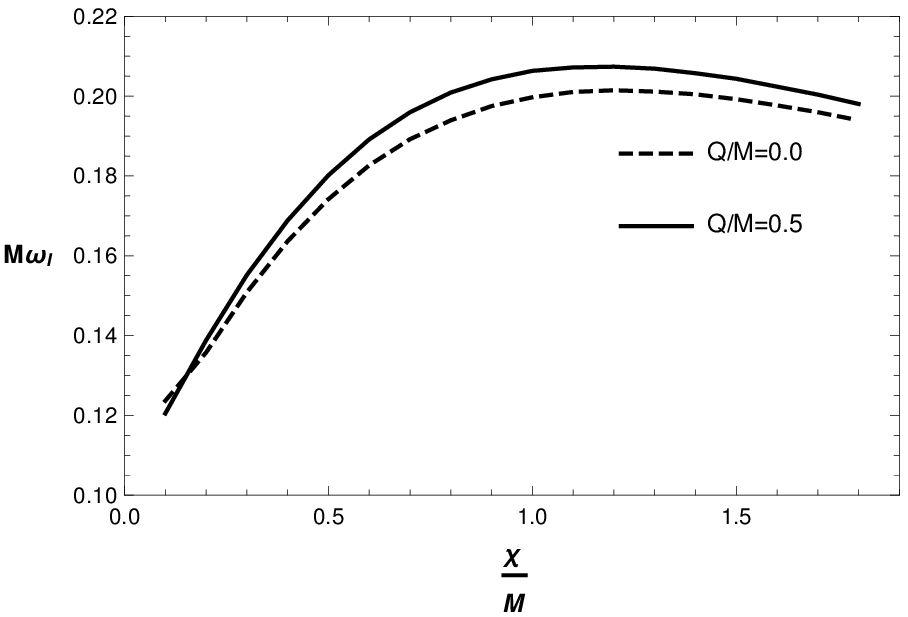}}
     \end{minipage}
  \caption{\footnotesize  Variation of the real ($M \omega_R$) and imaginary ($M \omega_I$) parts of the QNM frequencies with change in PFDM parameter $\frac{\chi}{M}$. The plots are for $l=1$ and $n=0$.}
  \label{5cb}
\end{figure}

\noindent The plots in Fig.\ref{5cb} show the variation of the real (left) and imaginary (right) parts of the frequency of quasinormal modes with increment in the PFDM parameter $\chi$. The plots are shown for the expression of $\omega$'s obtained from eq.(s) \eqref{44} and \eqref{45}. We find that both $\omega_R$ and $\omega_I$  increase with $\chi$ reaches a maximum and then decrease. The plots are shown for two different values of black hole charge $Q=$0.0 (black dashed) and 0.5 (black). We find that both frequencies are higher in presence of charge. The nature of variation of the frequencies with $\chi$ can be explained by the fact that $\omega \sim \frac{1}{Mass^b}$ where $b$ is a positive integer or fraction. The nature of the variation of mass of the black hole system is dictated by the PFDM parameter $\chi$ as discussed previously. That in turn dictates the nature of variation of QNM frequencies with $\chi$. The same is true for the case of charge $Q$ dependency of QNM.
\begin{figure}[H]
  \centering
  \begin{minipage}[b]{0.4\textwidth}
   {\includegraphics[width=\textwidth]{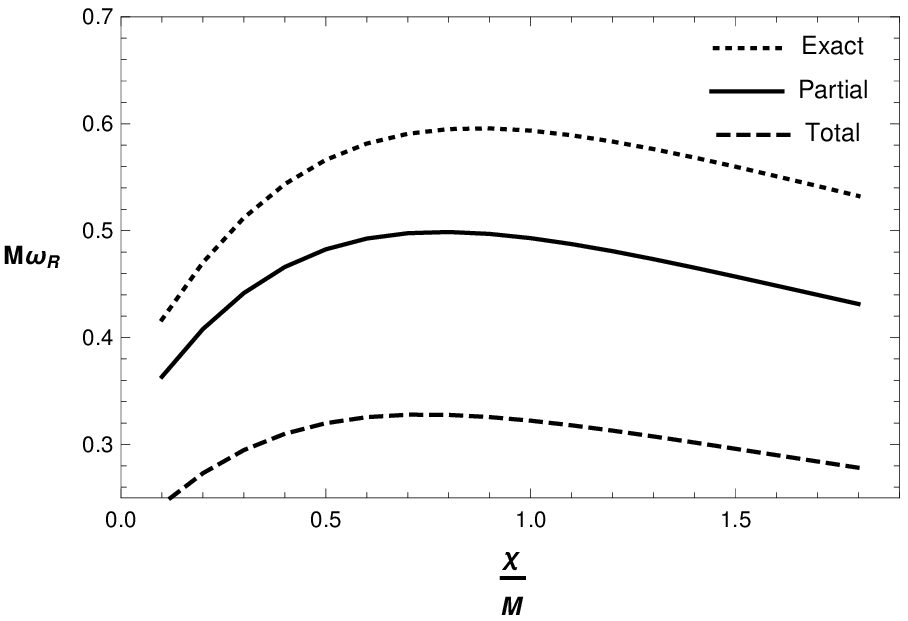}}
    \end{minipage}
  \hspace{1.0cm}
   \begin{minipage}[b]{0.4\textwidth}
    {\includegraphics[width=\textwidth]{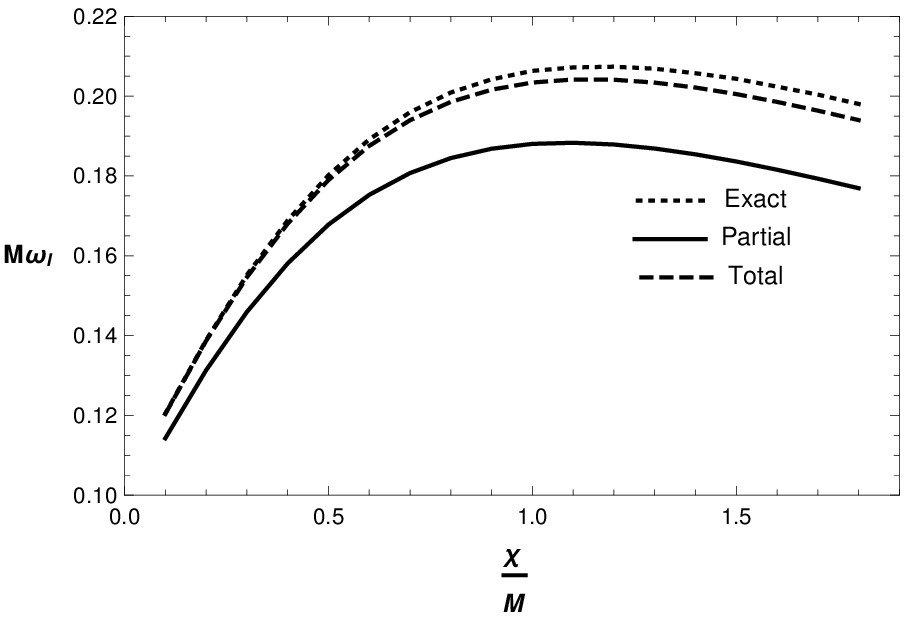}}
     \end{minipage}
  \caption{\footnotesize  Variation of the real ($M \omega_R$) and imaginary ($M \omega_I$) parts of the QNM frequencies with change in PFDM parameter $\frac{\chi}{M}$. The plots are for $l=1$ and $n=0$ with charge $\frac{Q}{M}=0.5$.}
  \label{5cb1}
\end{figure}

\noindent The plots in Fig. \ref{5cb1} show the variation of the real (left) and imaginary (right) parts of the $QNM$ frequencies with eq.\eqref{44} (Exact), eq.\eqref{52} (Partial) and eq.\eqref{54} (Total). We compare the different expressions for $l=1, n=0$ and found that there are certain differences in the results obtained. The results are obtained for the black hole charge $Q=0.5$.

\noindent $\omega_R$ obtained using the eikonal approximation $l>>1$ (Partial) have a much smaller value compared to the exact ones. Also, the values obtained using both $l>>1$ and $l>>n$ (Total) are smaller. Thus these approximations are valid only for large $l$ having a value around $l\sim 100$.  

\noindent Also, we find that in the case of $\omega_I$, the results are almost the same (Exact and Total) even for $l=1$ whereas the Partial one lies below them and can be matched only for higher values of $l$. 
\noindent The adjacent figure shows the variation of the quality factor ($QF$) of the perturbed black hole system. In general, the quality factor, as can be guessed by the name gives the quality of the resonating system. If $QF$ is large, then the system is said to be underdamped whereas if the $QF$ is small, then the system is said to be overdamped. Mathematically, it is defined as \cite{19}, \cite{52}
\begin{equation}
    Quality~~Factor = \frac{\omega_R}{2|\omega_I|}~.
\end{equation}
\begin{figure}[H]
\centering
\includegraphics[width=0.4\textwidth]{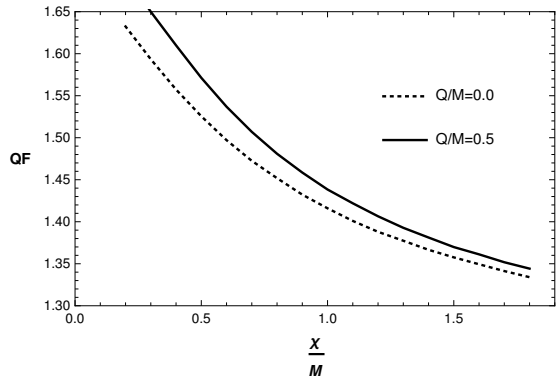}
 \caption{\footnotesize Variation of quality factor with change in $PFDM$ parameter $\frac{\chi}{M}$.}
 \label{g1}
\end{figure}

\noindent The plot in Figure \ref{g1} shows that with an increase in PFDM parameter $\chi$, the quality factor of the system reduces implying that dark matter over-damps the system. We have shown the plots for charges $Q=0.0$ (dotted black) and $Q=0.5$ (black). We find that the increase in charge $Q$ increases the $QF$ and thus we find that the presence of charge $Q$ under-damps the system. Hence the increase in charge $Q$ turns the system into a better oscillator.

\section{Shadow of charged black hole immersed in perfect fluid dark matter}\label{sec4}
Black hole shadow is formed by the null geodesics which are subject to the condition (eq.\eqref{31})
\begin{eqnarray}\label{91}
    2f(r_p)-rf'(r_p)=0 ~~\Rightarrow ~~2r_p ^2 -(6+\chi)r_p + 4Q^2 + 3 \chi r_pln\Big(\frac{r_p}{|\chi|}\Big)=0~.
\end{eqnarray}
Solving the above equation we get the value of photon sphere radius $r_p$ in terms of $\chi, Q$, that is $r_p = r_p \Big(\chi, Q\Big)$. $r_p$ has a maximum corresponding to some value of $\chi$ which can be yield using $\frac{\partial r_p}{\partial \chi}=0$. Using this condition, we get
\begin{equation}\label{92}
    (\chi_p)_m=\frac{1 +\sqrt{1 - \frac{4}{9} Q^2 \Big(2 + 3e^{-\frac{4}{3}}}\Big)}{1 + \frac{2}{3}e^{\frac{4}{3}}}~~;~~(r_p)_{m}=(\chi_p)_m e^{\frac{4}{3}}=\frac{1 +\sqrt{1 - \frac{4}{9} Q^2 \Big(2 + 3e^{-\frac{4}{3}}}\Big)}{\frac{2}{3} + e^{-\frac{4}{3}}}~.
\end{equation}
$(\chi_p)_m$ is the value of $\chi$ corresponding to the minimum of the photon sphere radius $(r_p)_m$. In the limit of $Q \to 0$, we get
\begin{equation}
    (\chi_p)_m=\frac{2}{1 + \frac{2}{3}e^{\frac{4}{3}}}~~;~~(r_p)_{m}=\frac{2}{\frac{2}{3} + e^{-\frac{4}{3}}}~.
\end{equation} 
\begin{figure}[H]
  \centering
  \begin{minipage}[b]{0.4\textwidth}
   {\includegraphics[width=\textwidth]{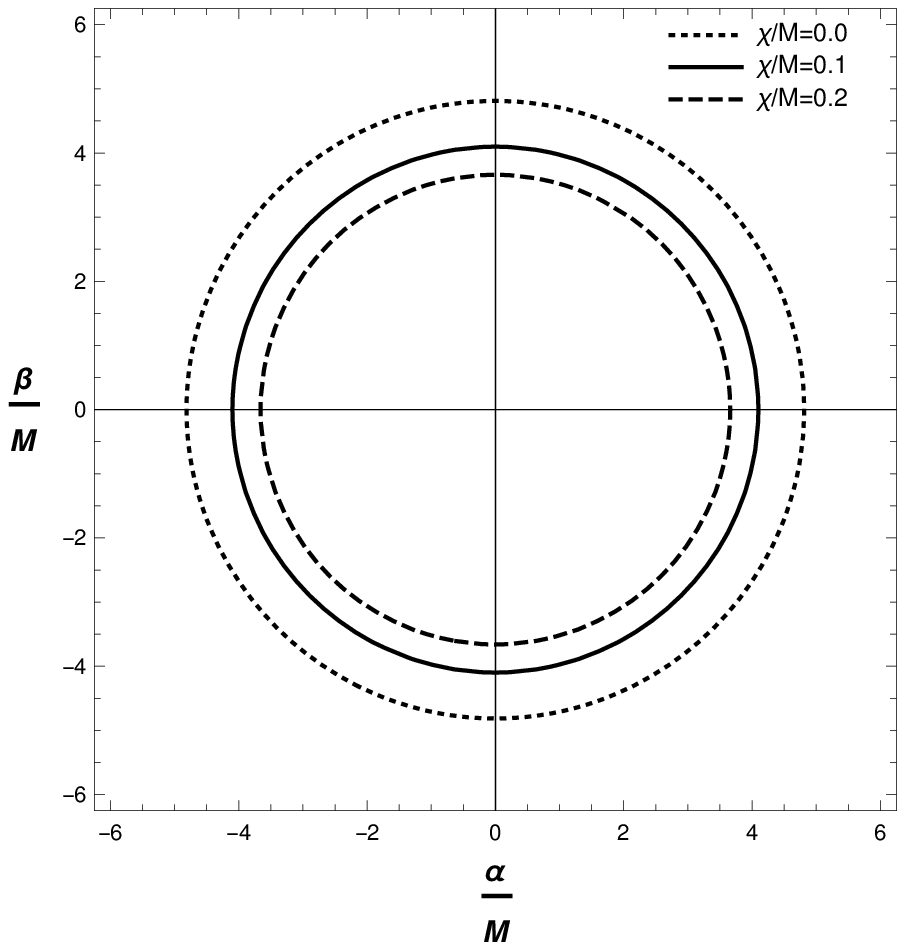}}
    \end{minipage}
  \hspace{1.0cm}
   \begin{minipage}[b]{0.4\textwidth}
    {\includegraphics[width=\textwidth]{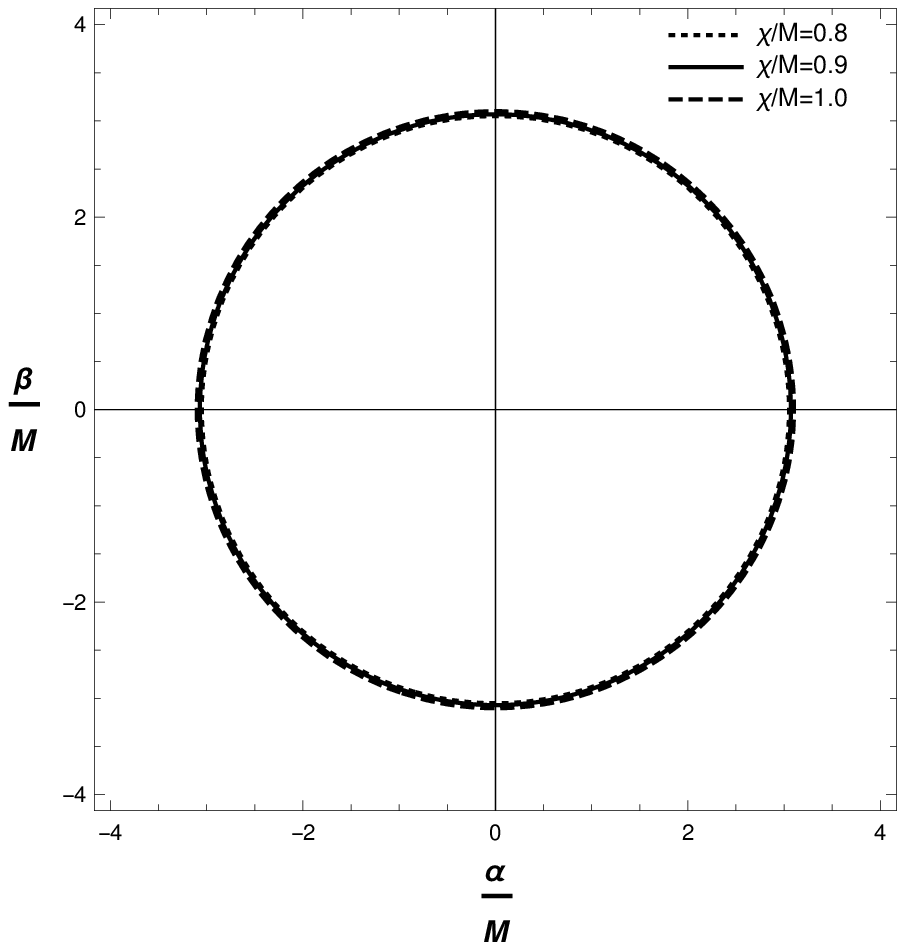}}
     \end{minipage}
  \caption{\footnotesize Contour-plots of black hole shadow with variation in $\frac{\chi}{M}$ for $\frac{\chi}{M} < \frac{\chi_c}{M}$ (left) and $\frac{\chi}{M} > \frac{\chi_c}{M}$ (right). The plots are shown for $\frac{Q}{M}=0.5$.}
  \label{5cbcb}
\end{figure}

\noindent To obtain eq.\eqref{92}, we first take the  derivative of eq.\eqref{91} with respect to $\chi$. Then we use the condition $\frac{\partial r_p}{\partial \chi}=0$ to obtain
\begin{equation}
   (r_p)_{m}=(\chi_p)_m e^{\frac{4}{3}}~. 
\end{equation}
Replacing $(r_p)_m$ in eq.\eqref{91}, we get $(\chi_p)_m$ and thereby $(r_p)_m$ as in eq.\eqref{92}.

\noindent The shadow radius is given as $R_s = \frac{r_p (\chi, Q)}{\sqrt{f(r_p (\chi, Q))}}$. This expression can be obtained using the null geodesics in the equatorial plane. The radial equation takes the form of eq.\eqref{5a} and the condition for unstable null circular geodesics gives eq.\eqref{31} and eq.\eqref{32}. 

\noindent Black hole shadow is formed by strong gravitational lensing of light rays in the vicinity of a black hole. Light from any background source comes near a black hole and travels in various trajectories. The light rays which move along the unstable circular geodesics can either fall into the black hole or travel to infinity upon encountering the slightest perturbation. For an observer placed in the equatorial plane $\theta_0 = \frac{\pi}{2}$ and at a distance $\bar{r}_0$ from the black hole, the angular shadow size can be calculated using $\tan \delta = \lim_{\Delta x \to 0} \frac{\Delta y}{\Delta x}$ \cite{77}. The expression can be rewritten in terms of geodesics. In the appropriate limit, we obtain $\tan \delta$ and thereby $\sin \delta$ as \cite{77}
\begin{equation}
    \tan \delta=\sqrt{\frac{f(r)}{r^2 \frac{E^2}{L^2}-f(r)}}\Bigg|_{r=\bar{r}_o}~~;~~\sin \delta= \sqrt{\frac{f(r)}{r^2}}\frac{L_p}{E_p}\Bigg|_{r=\bar{r}_o}=\sqrt{\frac{f(\bar{r}_0)}{\bar{r}_0 ^2}}\sqrt{\frac{r_p ^2}{f(r_p)}}~.
\end{equation}
For an observer positioned at a large distance from the black hole, the shadow radius takes the form
$R_s =\bar{r}_0 \tan \delta \approx \bar{r}_0 \sin \delta$~. In the limit of $\bar{r}_0 \to \infty$, we have $f(r) \to 1$ and the shadow radius $R_s$ takes the form $R_s = \frac{r_p}{\sqrt{f(r_p)}}$~.

\noindent The maximum value of $R_s$ corresponding to some values of $\chi$ can be found using the condition $\frac{\partial R_s}{\partial k}=0$. Using this condition, we get the values
\begin{equation}
    \chi_{rm}= \frac{3\Big(1+ \sqrt{1-\frac{8}{9}Q^2 (1+ e^{-1})}\Big)}{2(1+e)}~~;~~(R_s)_{rm} = \frac{\chi_{rm} e}{\sqrt{f(\chi_{rm}e)}}~.
\end{equation}
In the limit $Q \to 0$, we get \cite{23}, \cite{44}
\begin{equation}
    \chi_{rm}(Q=0)=\frac{3}{1+e}~.
\end{equation}
In Fig.\ref{5cbcb}, we have shown the contour plots of the black hole shadow for different values of PFDM parameter $\chi$. The plots are shown for black hole charge $Q=0.5$. The left plots are for $\chi < \chi_c$ and the right ones are for $\chi > \chi_c$. We find that the shadow size reduces with an increase in $\chi$ for  $\chi < \chi_c$ and increases with $\chi$ for  $\chi > \chi_c$. Also, we find that the effects are more pronounced for  $\chi < \chi_c$. The observation can be assigned to the fact that the shadow is formed by the light coming from the photon sphere. Thereby the shadow radius $R_s$ is dictated by the size of the photon sphere $r_p$ which in turn is dictated by the event horizon radius $r_+$ as mentioned previously. The variation of the event horizon with $\chi$ gets reflected on $r_p$ and thereby on $R_s$ and hence we observe such dependence of $R_s$ on PFDM parameter $\chi$. 

\begin{figure}[H]
  \centering
  \begin{minipage}[b]{0.4\textwidth}
   {\includegraphics[width=\textwidth]{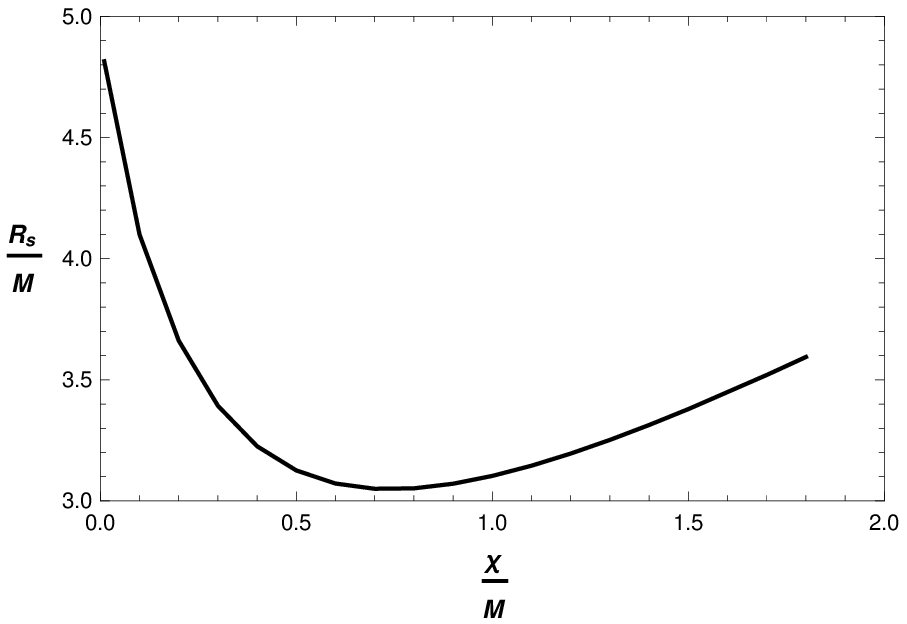}}
    \end{minipage}
  \hspace{1.0cm}
   \begin{minipage}[b]{0.4\textwidth}
    {\includegraphics[width=\textwidth]{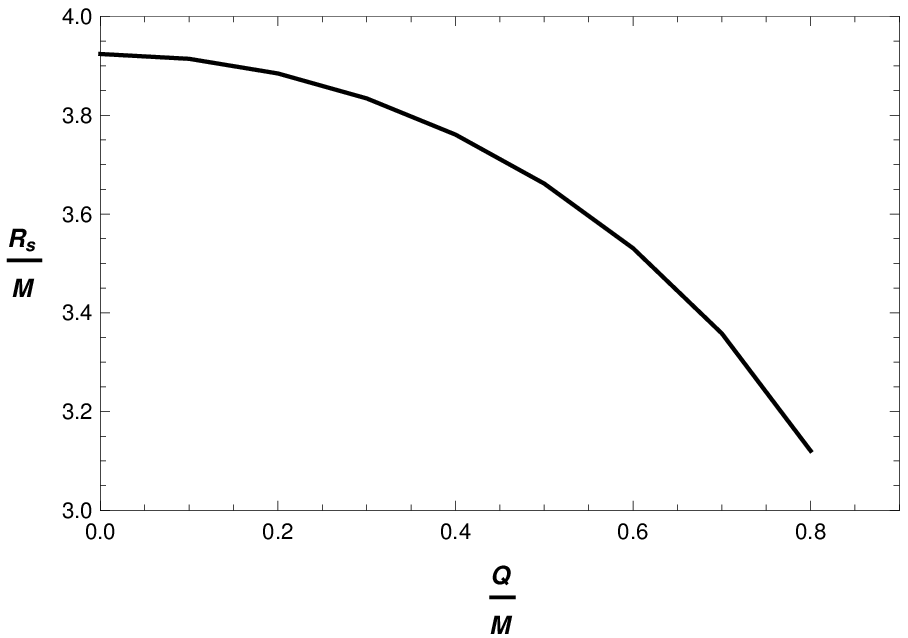}}
     \end{minipage}
  \caption{\footnotesize  Plots showing variation of shadow radius $R_s$ with $\chi$ (left) and $Q$ (right).}
  \label{5bbb}
\end{figure}

\noindent The left plot in Fig.\ref{5bbb} shows the variation of the shadow radius $R_s$ with the PFDM parameter $\chi$. We found that the shadow size initially decreases with $\chi$ reaching a minimum and then starts to increase. This observation can be explained by the dependence of $R_s$ on $r_+$ which in turn depends on $\chi$ as discussed above. On the other hand, the right plot shows the variation of shadow radius $R_s$ with the black hole charge $Q$. We find that the shadow radius decreases with an increase in black hole charge $Q$. This can also be explained by the dependence of $R_s$ on $r_+$ which decreases with an increase in charge $Q$ and that gets reflected in the shadow radius $R_s$.

\section{Summary and Conclusion}\label{sec5}
We summarize our findings now. We considered a charged black hole immersed in perfect fluid dark matter (PFDM). We studied the spacetime and found that the event horizon size $r_+$ depends on both PFDM parameter $\chi$ and charge $Q$. We find that the event horizon radius initially decreases with $\chi$ reaching a minimum at $\chi=\chi_c$ and then starts to increase. The expression for the \textit{minimum value} of $\chi_c$ and the corresponding value of $r_+$ are calculated and found to be dependent on charge $Q$ which is obtained for the first time in this work. We also find that the presence of PFDM parameter $\chi$ does not increase the number of horizons. The nature of variation of event horizon radius $r_+$ and thereby the black hole size with $\chi$ can be explained by the fact that the system is composed of two masses, one due to the black hole ($M$) and the other due to PFDM ($M_0$). Below the critical value $\chi_c$, the PFDM mass $M_0$ hinders the black hole mass $M$, and hence the effective mass of the system and thereby the size of event horizon $r_+$ decreases. But after $\chi_c$, the total mass of the system is dictated by the mass $M_0$ of the PFDM, thus $r_+$ increases. Also, we find that the size of the event horizon decreases with an increase in black hole charge $Q$. \\
\noindent Then we studied the variation of black hole temperature $T_h$  with the PFDM parameter $\chi$. We find that the temperature of the black hole initially increases with $\chi$ reaching a maximum and then starts to decrease. The reason being the temperature of the black hole is inversely dependent on the black hole's mass. Thus for $\chi < \chi_c$, the temperature of the system increases whereas the black hole mass decreases. On the other hand, for $\chi>\chi_c$, the temperature falls with an increase in the effective mass of the black hole system. Again, we also found that the black hole temperature $T_h$ increases with an increase in black hole charge $Q$. \\
\noindent Then we studied the timelike geodesics. The existence and finiteness of energy $E$ and angular momentum $L$ per unit mass of the massive particle constrain the values of circular orbit radius $r_0$ and PFDM parameter $\chi$ which can be represented as a parametric plot in $\chi-r_0$ plane. We find that the range of values of $r_0$ and $PFDM$ parameter $\chi$ increases with an increase in charge $Q$. Then by using the Lyapunov exponent ($\lambda$), we studied the existence of stable circular geodesics orbiting a black hole. The existence of stable geodesics puts a constraint on the possible values of $r_0$ and $\chi$ which we represent via a parametric plot in ($\chi- r_0$) plane. We find that the region of compatible values $\chi$ and $r_0$ increases with black hole charge $Q$. After that, we studied the critical exponent $\gamma$  and showed the parametric plot for  $\chi-r_0$ compatible with the detection of gravitational wave signals. We find that the compatible range for $r_0$ and $\chi$ increases with charge $Q$.

\noindent Then we studied the null geodesics where we find the dependence of photon sphere radius $r_p$ on $\chi$. The photon sphere radius $r_p$ initially decreases with $\chi$ reaching a minimum and then again starts to increase. We obtained the expression of the minimum value of $\chi$ and the corresponding one of $r_p$. The nature of variation of $r_p$ can be explained by the fact that the photon sphere radius indirectly gives the size of the event horizon radius $r_+$. Thus the dependence of $r_+$ on $\chi$ gets reflected in the nature of $r_p$. We also observed that the possibility of detection of gravitational waves in the case of null particles increases with an increase in charge $Q$.

\noindent Then we studied the effect of scalar field perturbation on the black hole background. Using the first-order WKB approximation, we calculated the real ($\omega_R$) and imaginary ($\omega_I$) parts of the quasinormal frequencies and their variation with the PFDM parameter $\chi$. We find that both $\omega_R$ and $\omega_I$ initially increase with $\chi$ reaching a maximum and then decrease. Also, we find that the QNM frequencies increase with the increase in charge $Q$. The observed variation of the QNM frequency with change in $\chi$ can be explained by considering $\omega \sim \frac{1}{(Mass)^b}$ (which is true for any massive oscillating system) where $b$ is a positive integer or fraction. The change in mass of the total system (black hole $+$ PFDM) is dictated by the PFDM parameter $\chi$ as discussed previously which in turn describes the nature of variation of QNM frequencies with $\chi$. Besides, we have also done a comparative study of the different expressions of QNM frequency. We find that the results for $\omega_R$ in the different cases are significantly varying for $l=1$ whereas the values of $\omega_I$ are quite close even for $l=1$. We also studied the quality factor ($QF$) of the oscillating black hole system and found that $QF$ gets reduced with an increase in PFDM parameter $\chi$ whereas it increases with the black hole charge $Q$. Thus increase in $\chi$ makes the system over-damped whereas the increment in charge turns the system into a better oscillator.  

\noindent Finally, we studied the black hole shadow and find the dependence of the shadow size and thereby the shadow radius $R_s$ on the PFDM parameter $\chi$. The shadow size reduces with an increase in $\chi$ reaches a minimum and then again starts to increase. On the other hand, the shadow size $R_s$ reduces with an increase in charge $Q$. This observation can be explained by the fact that shadow is formed by photons coming from photon spheres whose size dictates the black hole size or the size of the event horizon $r_+$. The dependence of $r_+$ on $\chi$ and $Q$ gets reflected on $r_p$ and thereby on $R_s$ and thus we obtain the above observation. Besides we have also obtained expressions for the minimum value of photon sphere radius $r_p$ and shadow radius $R_s$ corresponding to some value of PFDM parameter $\chi$ in the presence of charge $Q$ which was not done previously.

\section{Acknowledgements}
\noindent AD would like to acknowledge the support by SNBNCBS for the Senior Research Fellowship. ARC would like to acknowledge SNBNCBS for Senior Research Fellowship.

\section*{Appendix}\label{010a}
The equation whose solution we wish to obtain is
\begin{equation}\label{98}
     \frac{d^2 \psi(x)}{dx^2} + Q(x)\psi(x)=0~. 
\end{equation}
The solution of the above equation in the first order (where $Q(x)>0$) is given as \cite{73}
\begin{equation}
    \psi (x) = \frac{1}{Q(x)^{\frac{1}{4}}}exp\Big(\pm i \int_{x} ^{x_1}\sqrt{Q(x')}dx'\Big),~~~~~~~Region ~I
\end{equation}
\begin{equation}
    \psi (x) = \frac{1}{Q(x)^{\frac{1}{4}}}exp\Big(\pm i \int_{x_2} ^{x}\sqrt{Q(x')}dx'\Big),~~~~~~~~Region ~III
\end{equation}
where $x_1$ and $x_2$ are two turning points with $x_1 < x_2$. We want to obtain the solution for $\psi(x)$ in the region where $Q(x)<0$ for completeness. In the region $Q(x)<0$, $Q(x)$ can be approximated by a parabola. The approximation is valid if we assume the two turning points (where $Q(x)=0$) to be lying close to one another implying the region of $Q(x)<0$ is very small. In this region, $Q(x)$ can be expanded by using Taylor's expansion about the point of extrema of the potential. The point of extremum can be obtained by using $\frac{dQ(x)}{dx}\Big|_{x=x_0}=0$. In order to have continuous solutions, we obtain the asymptotic solutions of the region $Q(x)<0$ and match them to the ones in $Q(x)>0$. 
\begin{figure}[h]
\centering
\includegraphics[width=0.4\textwidth]{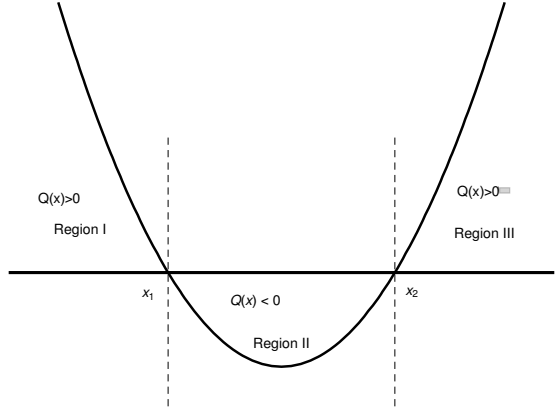}
\caption{\footnotesize The function $Q(x)$ in different regions~.}
\label{g}
\end{figure}
The Taylor expansion gives $Q(x)$ as 
\begin{equation}\label{99}
    Q(x)=Q(x_0)+\frac{1}{2}Q''(x)\Big|_{x=x_0}(x-x_0)^2 + O(x-x_0)^3,~~~~Region ~II~.
\end{equation}
To find the solution in the region $Q(x)<0$, we replace $Q(x)$ from eq.\eqref{99} in eq.\eqref{98} upto $(x-x_0)^2$ which gives
\begin{equation}
     \frac{d^2 \psi(x)}{dx^2} + \Big(Q(x_0)+\frac{1}{2}Q''(x_0)(x-x_0)^2\Big)\psi(x)=0 
\end{equation}
In order to obtain the solution, we modify the equation in such a way that the equation can have an exact solution. The exact solution can be represented in terms of parabolic cylinder functions \cite{78}. We need to use
\begin{equation}
    \kappa \equiv \frac{1}{2}Q''(x_0)~~;~~\eta \equiv (4 \kappa)^{\frac{1}{4}}e^{i\frac{\pi}{4}}(x-x_0)~~;~~\mu + \frac{1}{2}\equiv\frac{-iQ(x_0)}{\sqrt{2Q''(x_0)}}~.
\end{equation}
which gives the modified equation in the form
\begin{equation}
    \frac{d^2 \psi (\eta)}{d\eta^2} + \Big(\mu + \frac{1}{2}- \frac{1}{4}\eta^2\Big)\psi (\eta) = 0~.
\end{equation}
The solution of the equation can be obtained in terms of parabolic cylinder function $D_{\mu}(\eta)$. Also, $D_{\mu}(-\eta)$ and $D_{-\mu-1}(-i\eta)$ are also solutions of the above equation. So they must be linearly dependent. So $D_{\mu}$ can be written as a linear combination of $D_{\mu}(-\eta)$ and $D_{-\mu-1}(-i\eta)$. Thus,
\begin{equation}
    \psi (\eta)=D_{\mu}(\eta)= A_1 D_{\mu}(-\eta) + A_2 D_{-\mu-1}(-i\eta) 
\end{equation}
where $A_1$ and $A_2$ are coefficients. Using $D_{\mu} (0)$ and $D'_{\mu} (0)$, the coefficients $A_1$ and $A_2$ can be determined and the final solution takes the form \cite{78}
\begin{equation}
 \psi (\eta)= e^{i\mu \pi} D_{\mu}(-\eta) + \frac{\sqrt{2 \pi}}{\Gamma(-\eta)} D_{-\mu-1}(-i\eta)~.    
\end{equation}
In the asymptotic limit, the solution $\psi(\eta)$ can be explicitly expressed as
\begin{gather}
    \psi(\eta) \sim \eta^{\mu}e^{-\frac{\eta^2}{4}}\Bigg(1-\frac{\mu(\mu-1)}{2\eta^2} + \frac{\mu(\mu-1)(\mu-2)(\mu-3)}{2.4\eta^4}_......\Bigg)\\ \nonumber
    - \frac{\sqrt{2\pi}}{\Gamma{(-\mu)}}e^{i\mu \pi}\eta^{-\mu-1}e^{\frac{\eta^2}{4}}\Bigg(1+\frac{(\mu+1)(\mu+2)}{2\eta^2} + \frac{(\mu+1)(\mu+2)(\mu+3)(\mu+4)}{2.4\eta^4}_......\Bigg) = \psi_{I} + \psi_{II}~.
\end{gather}
We find that the first term goes as 
\begin{equation}
    \psi_{I} \sim e^{-\frac{\eta^2}{4}} = e^{-i\sqrt{\frac{\kappa}{4}}(x-x_0)^2}
\end{equation}
and the second term goes as
\begin{equation}
   \psi_{II} \sim e^{\frac{\eta^2}{4}} = e^{i\sqrt{\frac{\kappa}{4}}(x-x_0)^2}~.
\end{equation}
In the asymptotic limits, we require the solution to be of the form $\psi \sim e^{-ia(x-x_0)^2}$, where $a$ is some function of $Q(x)$ which in the asymptotic limit is a constant. So $\psi_{II}$ term must be zero. The only way out is to set $\Gamma(-\mu)=\infty$ which implies $\mu$ to be any positive integer $n$ ($n =0,1,2,3,....$). Thus, we obtain
\begin{equation}
   n + \frac{1}{2}\equiv\frac{-iQ(x_0)}{\sqrt{2Q''(x_0)}} 
\end{equation}
which we designate as the condition for the determination of quasinormal frequencies or the quasi-normal mode ($QNM$) condition.

\end{document}